\documentstyle[psfig]{mn}

\newif\ifAMStwofonts

\newcommand{\lapp}{\mbox{\raisebox{-0.3em}{$\stackrel{\textstyle <}{\sim}$}}}

\title[The double-double radio galaxy J1453+3308]
      {Spectral ageing analysis of the double-double radio galaxy J1453+3308}
\author[C. Konar et al.]
       {C. Konar$^1$$\thanks{E-mail: sckonar@ncra.tifr.res.in (CK); djs@ncra.tifr.res.in (DJS); 
        jamrozy@oa.uj.edu.pl (MJ); machalsk@oa.uj.edu.pl (JM)}$, D.J. Saikia$^1$, M. Jamrozy$^2$ 
        and J. Machalski$^2$ \\
$^1$ National Centre for Radio Astrophysics, TIFR, Pune University Campus, Post Bag 3, Pune 411 007, India \\
$^2$ Obserwatorium Astronomiczne, Uniwersytet Jagiello\'nski, ul. Orla 171, 30244 Krak\'ow, Poland }
\date{Accepted.    Received }

\pagerange{\pageref{firstpage}--\pageref{lastpage}}
\pubyear{  }

\begin{document}

\maketitle

\label{firstpage}

\begin{abstract}
We present new radio observations at frequencies ranging from 240 to 4860 MHz
of the well-known, double-double radio galaxy (DDRG), J1453+3308, using both the 
Giant Metrewave Radio Telescope (GMRT) and the Very Large Array (VLA). These 
observations enable us to determine the spectra of the inner and outer lobes over
a large frequency range and demonstrate that while the spectrum of the outer lobes 
exhibits significant curvature, that of the inner lobes appears practically straight.
The break frequency, and hence the inferred synchrotron age of the outer structure,
determined from 16$-$arcsec strips transverse to the source axis, increases with
distance from the heads of the lobes. The maximum spectral ages for the northern and southern
lobes are $\sim$47 and 58 Myr respectively. Because of the difference in the lengths
of the lobes these ages imply a mean separation velocity of the heads of the lobes
from the emitting plasma of 0.036c for both the northern and southern lobes. 
The synchrotron age of the inner double is about 2 Myr which implies an 
advance velocity of $\sim$0.1c,
but these values have large uncertainties because the spectrum is practically straight.
\end{abstract}


\section{Introduction}
For radio galaxies and quasars, an interesting way of probing their history 
is via the structural and spectral information of the lobes of extended 
radio emission. Such studies have provided useful insights in understanding sources
which exhibit precession or changes in the ejection axis, effects of motion of the 
parent galaxy, backflows from hotspots as well as X-shaped sources and major 
interruptions of jet activity.  A
striking example of episodic jet activity is when a new pair of radio lobes
is seen closer to the nucleus before the `old' and more distant radio
lobes have faded (e.g. Subrahmanyan, Saripalli \& Hunstead 1996; Lara et al. 1999).
Such sources have been christened as `double-double' radio galaxies
(DDRGs) by Schoenmakers et al. (2000a, hereinafter referred to as S2000a). 
They proposed a relatively general definition of a DDRG as a double-double radio
galaxy consisting of a pair of double radio sources with a common centre. S2000a also
suggested that the two lobes of the inner double should have an edge-brightened
radio morphology to distinguish it from knots in a jet. In such sources the newly-formed
jets  may propagate outwards through the cocoon formed by the earlier cycle of activity
rather than the general intergalactic or intracluster medium after traversing through the 
interstellar medium of the host galaxy. Approximately a dozen or so of such DDRGs are known
in the literature (Saikia, Konar \& Kulkarni 2006, and references therein).

Kaiser, Schoenmakers \& R\"{o}ttgering (2000, hereinafter referred to as K2000) presented
an analytical model predicting how the new jets give rise to an inner source structure
within the region of the old, outer cocoon. Assuming values of a number of free model
parameters, e.g. the density of the external environment surrounding the old jets, the
exponent of the density distribution as a function of the radial distance from the radio
core, the initial energy distribution in the shock caused by the jet, they predicted
a lower limit for the power of the old jets, and dynamical ages of the outer (older), $t_{out}$,
and inner (younger), $t_{inn}$, double structures of a few selected DDRGs. In particular,
for the source J1453+3308 they found $t_{out}=215$ Myr and $t_{inn}=1.6$ Myr.

In this paper we investigate the radiative ages of particles emitting in different
parts of the old cocoon (outer structure) of J1453+3308, as well as in the inner lobes of
this DDRG, and compare these with the above dynamical ages. The already available data
indicate that the lobes of the outer double are
separated by 336 arcsec corresponding to  1297 kpc, while the inner double has a separation
of 41 arcsec corresponding to 159 kpc (H$_\circ$=71 km s$^{-1}$ Mpc$^{-1}$, $\Omega_m$=0.27, 
$\Omega_\Lambda$=0.73, Spergel et al. 2003). The 1.4-GHz radio luminosities of the outer and
inner doubles are 7.94$\times$10$^{25}$ W Hz$^{-1}$ and 
5.88$\times$10$^{24}$ W Hz$^{-1}$ respectively. The luminosity of the outer double is above
the FRI/FRII break, while that of the inner double is below it although it has an
edge-brightened structure.  

To achieve the above objectives we made new radio maps of J1453+3308
at a number of frequencies over a large range from 240 to 4860 MHz. 
These observations have been made with an
angular resolution high enough to image the
lobes with at least 6 resolution elements along their axes. The new observations 
and data reduction are
described in Section~2. The observational results, such as the radio maps showing the
source structure, spectra and polarisation parameters 
are presented in Section~3. The standard spectral-ageing analysis for the outer and inner
structures is described and the results presented in Section~4, while the concluding
remarks are given in Section~5.

\section{Observations and data reduction}
The analysis presented in this paper is based on radio observations
recently conducted with the GMRT and VLA, as well as on VLA archival data.
The observing log for both the GMRT and VLA observations is listed in
Table~1 which is arranged as follows. Columns 1 and 2 show the name of the
telescope, and the array configuration for the VLA observations;
columns 3 and 4 show the frequency and bandwidth used in making the images;
column 5: the primary beamwidth in arcmin; column 6: dates of the observations.
The phase centre for all the observations was near the core of the radio
galaxy.

\subsection{GMRT observations}
The observations were made in the standard manner, with  each observation
of the target-source interspersed with observations of 3C286 which was
used as a phase calibrator as well as flux density and bandpass calibrator.
At each frequency the source was observed in a
full-synthesis run of approximately 9 hours including calibration overheads.
The rms noise in the resulting images range from about 1 mJy\,beam$^{-1}$ at
240 MHz to about 0.06 mJy\,beam$^{-1}$ at 1287 MHz.
Details  about the array can be found at the GMRT website at
{\tt http://www.gmrt.ncra.tifr.res.in}. The data collected were calibrated 
and reduced in the standard way using the NRAO {\tt AIPS} software package.
The flux densities at the different frequencies are based on the scale of
Baars et al. (1977). 

\begin{table}
\caption{Observing log. }
\begin{tabular}{l c r c c c }
\hline
Teles-    & Array  & Obs.   & Band- & Primary &  Obs. Date  \\
cope      & Conf.  & Freq.  & width &  beam   &             \\
          &        & MHz    & MHz   & arcmin  &             \\
\hline
GMRT      &        & 240    & 4.5   &  114    & 2005 Mar 17   \\
GMRT      &        & 334    & 12.5  &   81    & 2004 Dec 25   \\
GMRT      &        & 605    & 12.5  &   43    & 2005 Mar 17   \\
GMRT      &        & 1287   & 12.5  &   26    & 2005 Dec 22   \\
VLA$^\ast$&    A   & 1365   &  25   &   30    & 2000 Oct 21    \\
VLA$^\ast$&    A   & 4860   &  50   &    9    & 2000 Oct 22    \\
VLA$^\ast$&    B   & 4860   &  50   &    9    & 2001 Apr 02    \\
VLA$^\ast$&   AB   & 4860   &  100  &    9    & 1998 Jul 29    \\
VLA       &  CnD   & 4860   &  50   &    9    & 2005 Oct 31    \\
VLA$^\ast$&    A   & 8460   &   50  &   5.4   & 2000 Oct 22    \\
VLA$^\ast$&    B   & 8460   &  50   &   5.4   & 2001 Apr 02    \\
\hline
\end{tabular}

$^\ast$: VLA archival data.
\end{table}

\subsection{New and archival VLA observations}
The source was observed with the CnD array at a
frequency of 4860 MHz to image the outer lobes and determine
their spectra by comparing with the low-frequency GMRT images. 
The integration time was about $6\times20$ min, which
allowed us to reach an rms noise value of about 0.05 mJy\,beam$^{-1}$. The
interferometric phases were calibrated every 20 min with the phase calibrator
J1416+347. The source 3C286 was used as the primary flux density
and polarisation calibrator. For the image produced from this data
set correction for the primary beam pattern has been done. 

As in the case of the GMRT data, the VLA data were edited and reduced using the
{\tt AIPS} package. The polarisation data reduction was done and the maps of the
Stokes parameters $I$, $Q$ and $U$ were obtained using the procedures applied
in the analysis of polarisation properties in a larger set of giant radio
galaxies (Machalski et al. 2006).

To study the inner double-lobed structure of J1453+3308, derive its spectrum and
estimate the radiative age, as well as to compare the resulting age with that
of the outer lobes of the giant-sized structure, we supplemented our GMRT observations 
with VLA archival data. These observations which 
were made in the snap-shot mode in the L (1365 MHz),
C (4860 MHz) and X (8460 MHz) bands, also enabled us to study the spectrum and variability
of the radio core.  All flux densities are on the Baars et al. (1977) scale.  

\section{Observational results}
\subsection{Structure}
The images of the entire source using the GMRT and the VLA are presented in
Figs. 1 and 2, while the observational parameters and some of the observed
properties are presented in Table 2 which is arranged as follows.
Column 1: frequency  of observations in MHz, with the letter G or V representing
either GMRT or VLA observations; columns 2$-$4: the major and minor axes of the restoring
beam in arcsec and its position angle (PA) in degrees; column 5: the rms noise in
mJy\,beam$^{-1}$; column 6: the integrated flux density of the source in mJy estimated
by specifying an area around the source; column 7, 10, 13 and 16: component designation
where N1 and S1 indicate the northern and southern components of the outer double, N2 and
S2 the northern and southern components of the inner double; columns 8 and 9, 11 and 12,
14 and 15, 17 and 18: the peak and total flux densities of the components in
mJy\,beam$^{-1}$ and mJy respectively.  The flux densities have been estimated
by specifying an area around each component. The error in the flux density is approximately
15 per cent at 240 MHz and 7 per cent at the higher frequencies. 

\begin{figure*}
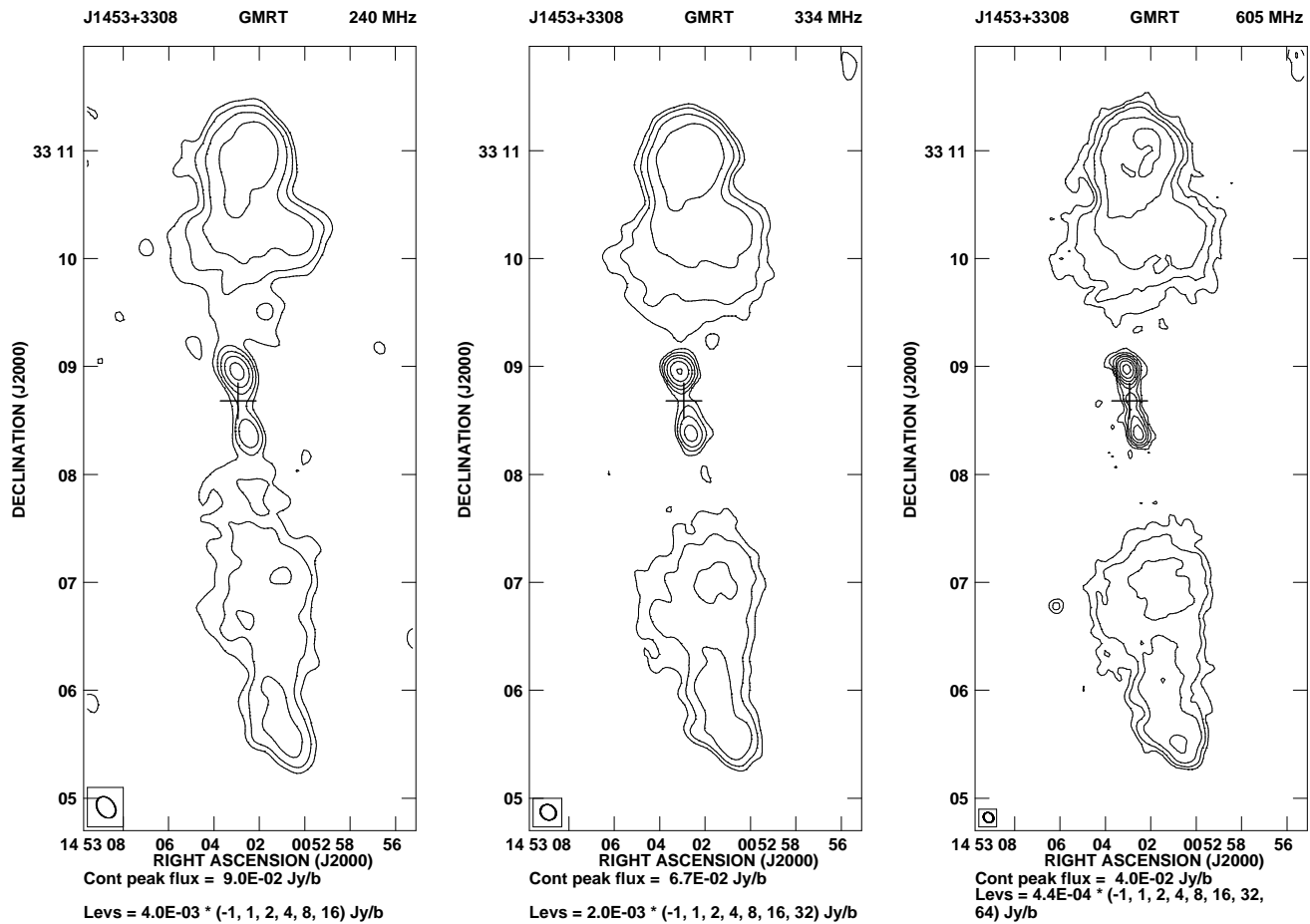

\hbox{
 \psfig{file=J1453+3309_239.PS,width=2.3in,angle=0}
 \psfig{file=J1453+3309_P.PS,width=2.3in,angle=0}
 \psfig{file=J1453+3309_610.PS,width=2.3in,angle=0}
}
\caption[]{The GMRT images of J1453+3308. Left panel: Image at 240 MHz; middle panel: the
334-MHz image; right panel: the image at 605 MHz.
In all the images presented in this paper the restoring beam is indicated by an ellipse,
and the $+$ sign indicates the position of the optical galaxy. The peak brightness and contour
levels are given below each image.}
\end{figure*}

\begin{figure*}
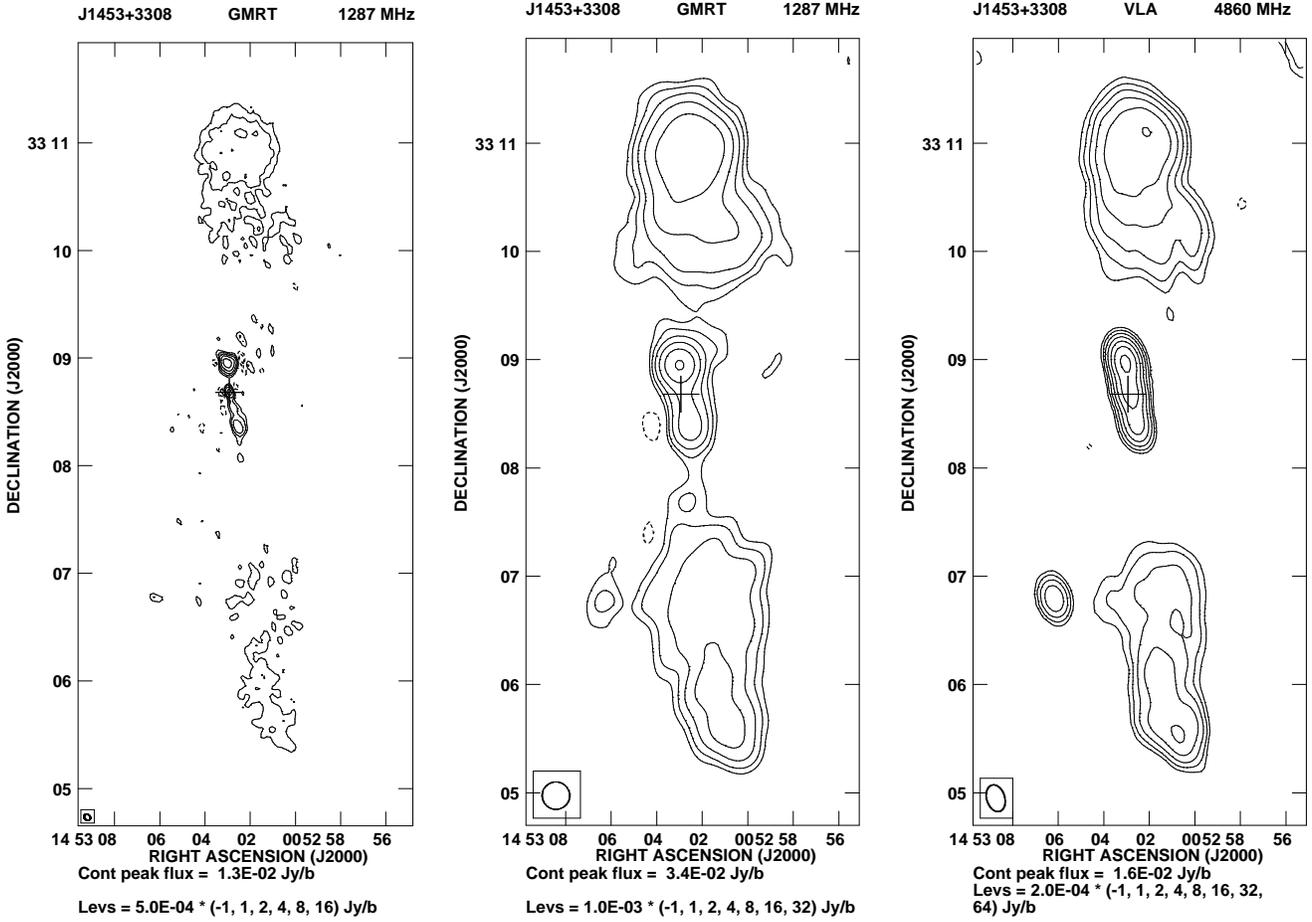

\hbox{
   \psfig{file=G.J1453_L.hires_1.ps,width=2.3in,angle=0}
   \psfig{file=J1453+3309_L_lores.PS,width=2.3in,angle=0}
   \psfig{file=J1453+3309_C.PS,width=2.3in,angle=0}
}
\caption[]{Left and middle panels: GMRT images of J1453+3308 at 1287 MHz with the full resolution
(cf. Table 2) and smoothed to a resolution of 15.2$\times$15.2 arcsec$^2$ respectively.
Right panel: VLA image at 4860 MHz. A weak ($\sim$2.5 mJy) unresolved source at R.A.:
14 53 06.20 and Decl.: $+$33 06 47.8 is unrelated to J1453+3308}
\end{figure*}

All the images of the source show the well-known pair of doubles which are misaligned by
$\sim$7.5$^\circ$. The total luminosity of the outer double is in the FRII category
although it has no prominent hotspot. On the other hand, the total luminosity of
the inner double is in the FRI class although it has an edge-brightened structure 
characteristic of FRII sources. Such sources could provide useful insights towards understanding
the FRI-FRII dichotomy and it would be useful to enquire whether the observed properties of the
inner double might be due to evolutionary effects such as the emissivity increasing with time  
or due to differences in the environment or the central engine.  

The high-resolution GMRT L-band image
shows the radio core in addition to the lobes of radio emission. The structure of the inner double
is better seen in the VLA A-array image at 1365 MHz and the B-array image at 4860 MHz,
made from the archival data (Fig. 4).

The ratio of the separations of the  southern lobe from the core to that of the northern one
is 1.34 and 1.06 for the outer and inner doubles, respectively. The outer double being more
asymmetric is possibly due to asymmetries in the external environment on the scale of the outer
double. The flux densities of the lobes are higher on the northern side, the ratio of the flux
density of the southern component to that of the northern one for the outer and inner doubles
are 0.57$\pm$0.06 and 0.44$\pm$0.05 respectively. This suggests an intrinsic asymmetry
on scales ranging from the inner double to that of the outer one.

\begin{table*}
\caption{ The observational parameters and flux densities of the outer (N1 and S1) and
inner (N2 and S2) lobes of J1453+3308}.
\begin{tabular}{l rrr r r l rr l rr l rr l rr}
\hline
Freq.       & \multicolumn{3}{c}{Beam size}                    & rms      & S$_I$   & Cp  & S$_p$  & S$_t$  & Cp   & S$_p$ & S$_t$ & Cp  & S$_p$   & S$_t$   &  Cp  & S$_p$ & S$_t$   \\
                                                                                                                          
            MHz         & $^{\prime\prime}$ & $^{\prime\prime}$ & $^\circ$ &    mJy   & mJy     &     & mJy    & mJy    &      & mJy   & mJy   &     & mJy     & mJy   &   & mJy & mJy    \\
                        &                   &                   &          &  /b      &         &     & /b     &        &      & /b    &       &     & /b      &        &   &  /b &       \\
(1) & (2) & (3) & (4) & (5) & (6) & (7) & (8) & (9) & (10) & (11) & (12) & (13) & (14) & (15) & (16) & (17) & (18 ) \\
\hline
  G240       & 12.8    &  9.4       &  35                       &   1.07   &  1843   & N1  &  53    &1055    & N2   &  92   & 123   & S2  &  31     &  48  & S1 & 22  & 602     \\

  G334       &  9.2    &  8.1       &  48                       &   0.52   &  1599   & N1  &  29    & 902    & N2   &  66   &  96   & S2  &  23     &  47  & S1 & 14  & 563     \\
                                                                                                                          
  G605       &  6.0    &  5.0       &  43                       &   0.11   &  1078   & N1  &  8.3   & 606    & N2   &  42   &  69   & S2  &  13     &  29  & S1 & 4.2 & 370     \\

  G1287      &  3.8    &   3.1      &  47                       &   0.06   &   504   & N1  &  2.4   & 272    & N2   &  13   &  37   & S2  & 3.8     &  17  & S1 & 1.2 & 164     \\
                                                                                                                          
 V1400       & 45.0    & 45.0       &                           &   0.45   &   460   & N1  &  121   & 258    &      &       &       &     &         &      & S1 & 46  & 145    \\
                                                                                                                          
 V4860       & 15.2    & 10.0       &  17                       &   0.05   &   136   & N1  &  6.5   &  71    & N2+S2&  16   &  32   &     &         &      & S1 & 3.6 &  32    \\
                                                                                                                          
 V1365       &  1.5    &   1.3      &   9                       &   0.02   &         &     &        &        & N2   &  6.1  &  38   & S2  & 2.3     &  14  &    &     &         \\
                                                                                                                          
 V4860       &  1.3    &   1.2      &  44                       &   0.02   &         &     &        &        & N2   &  2.2  &  16   & S2  & 0.8     &  4.9 &    &     &         \\
                                                                                                                          
 V8460       &  0.7    &   0.7      &   9                       &   0.01   &         &     &        &        & N2   &  0.5  &  9.5  & S2  & 0.2     &  2.6 &    &     &         \\
                                                                                                                          
\hline
\end{tabular}
\end{table*}

\begin{table*}
\caption{Flux densities of the total source, and its inner and outer lobes used for
fitting a spectral shape.}
\begin{tabular}{r c r r c r r c r r c}
\hline
Freqency  &Component& $S_{t}$ &  Error & Component & $S_{t}$ & Error & Component & $S_{t}$ & Error & Reference\\
 MHz   &     &  mJy    &  mJy  &       & mJy     & mJy  &       & mJy     & mJy      \\
(1)    &  (2)    &  (3)    & (4)  &  (5)   &  (6)    & (7)  & (8)   & (9)    & (10)  & (11) \\
\hline
 74    & total   &  2530   &  320  &&&&&&&   VLSS     \\
 151   & total   &  2387   &  135  &       &         &      & outer & 2165    & 110  &   1 \\
 178   & total   &  2220   &  220  &       &         &      & outer & 2020    & 200  &   2 \\
 240   & total   &  1843   &  260  & inner & 176     & 25   & outer & 1667    & 250  &   3 \\
 325   & total   &  1510   &  150  &       &         &      & outer & 1365    & 140  &   4 \\
 334   & total   &  1599   &  113  & inner & 143     & 10   & outer & 1456    & 112  &   3 \\
 408   & total   &  1330   &  125  &       &         &      &       &         &      &   5 \\
  605  & total   &  1078   &   76  & inner & 108     &  8   & outer &  970    &  75  &   3 \\
 1287  & total   &   504   &   36  & inner &  62     &  9   & outer &  442    &  34  &   3 \\
 1400  & total   &   460   &   33  &       &         &      & outer &  426    &  26  &   4,6 \\
 4860  & total   &   136   &   10  & inner &  32     &  3   & outer &  104    &   8  &   3 \\   
 4800  & total   &   121   &   18  &       &         &      &       &         &      &   4 \\
10500  & total   &    70   &    3  & inner &  20     &  4   &       &         &      &   4 \\
\hline
\end{tabular}
\begin{flushleft}

References: VLSS: VLA Low-frequency Sky Survey;
1: Riley et al. 1999;
2: Pilkington \& Scott (1965);
3: Present  paper;
4: Schoenmakers et al. (2000b)
5: Colla et al. (1970);
6: Schoenmakers et al. (2000a). \\
\end{flushleft}
\end{table*}

\subsection{Spectra}
The integrated  flux densities of the total source, as well as of the inner and outer doubles,
are given in Table~3. All columns are self explanatory, with `outer' and `inner' denoting the
outer lobes and the inner double.  All the flux densities quoted here are consistent
with the  scale of Baars et al. (1977). 
The flux densities of the outer lobes at 151, 178 and 325 MHz
(column 9) are derived subtracting flux densities expected for the inner lobes at these
frequencies from the total flux densities given in column 3. The former ones are simply
determined fitting the form log\,$S$=a + b\,log$\nu$ + c\,exp($-$log$\nu$) to the measured
flux density values in column 6.

The integrated spectrum is presented in Fig. 3. Clearly there is a
distinct steepening in the integrated spectrum at $\sim$300 MHz.  Our
low-frequency measurements show a significant flattening of the spectrum below $\sim$300 MHz,
consistent with the VLA Low-frequency Sky Survey (VLSS; {\tt http://lwa.nrl.navy.mil/VLSS}) 
flux density.  Since the low-frequency flattening is unlikely to be due
to either synchrotron self-absorption or thermal absorption, the change in curvature must be
due to spectral ageing with the low-frequency spectrum probably representing the injection
spectrum of the relativistic particles. Further analysis of this aspect is described in
Section~5.

\begin{figure}
\vbox{
   \psfig{file=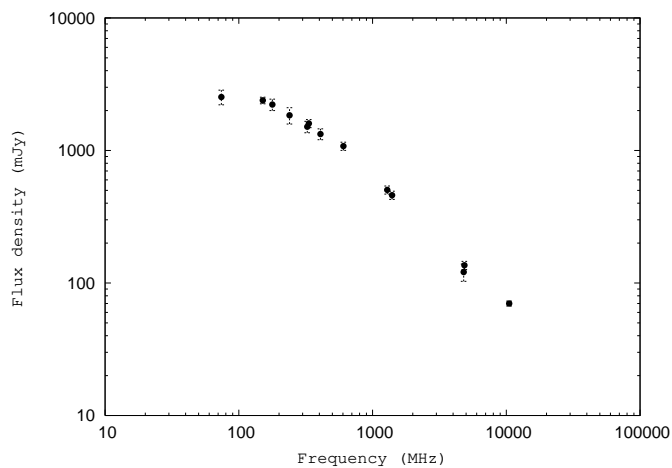,width=3.5in,angle=-90}
}
\caption[]{The integrated spectrum of J1453+3308.}
\end{figure}

\begin{figure}
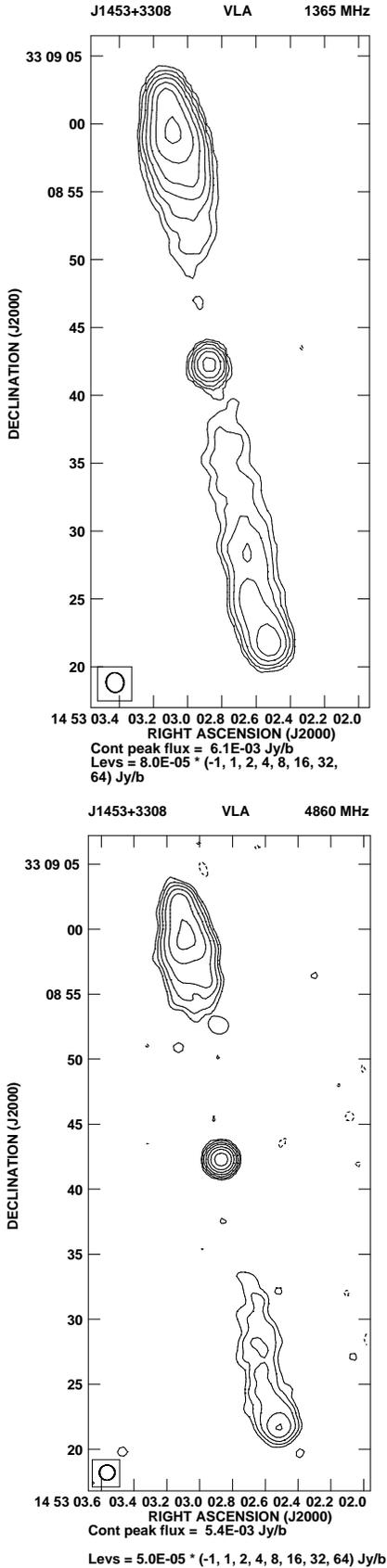

\vbox{
   \psfig{file=V_A.J1453L.inn_2.ps,width=2.2in,angle=0}
   \psfig{file=1450+333-INN-C.PS,width=2.2in,angle=0}
}
\caption[]{Images of the inner structure of J1453+3308
at 1365 MHz (upper panel) and 4860 MHz (lower panel) made from archival VLA data.}
\end{figure}

\subsection{The radio core}
The J2000.0 position of the radio core estimated from our high-resolution images is 
RA: 14 53 02.87 and DEC: 33 08 42.3, which is $\sim$2 arcsec away from the 
position of the optical galaxy (RA: 14 53 02.93 and DEC: 33 08 40.8) listed by 
Schoenmakers (1999).  The radio core has a mildly inverted spectrum
(Fig. 5) with only marginal evidence of variability at 4860 MHz but significant 
variability at 8460 MHz over a similar time scale.

\begin{table}
\caption{Flux densities of the radio core}
\begin{tabular}{c c r r }
\hline
Telescope    &  Date of obs. & Freq. &  Flux   \\
             &               &       &  density  \\
             &               &  MHz  &   mJy     \\
\hline
 GMRT        & 2005 Mar 17   & 605   &  3.1   \\
 VLA-A       & 2000 Oct 21   & 1365  &  4.1   \\
 VLA-B       & 2001 Apr 02   & 4860  &  5.5   \\
 VLA-A       & 2000 Oct 22   & 4860  &  5.5   \\
 VLA-AB      & 1998 Jul 29   & 4860  &  4.6   \\
 VLA-B       & 2001 Apr 02   & 8460  &  6.2   \\
 VLA-A       & 2000 Oct 22   & 8460  &  3.6   \\
\hline
\end{tabular}
\begin{flushleft}
The core flux density at 605 MHz is the peak
value, due to contamination by the extended flux
density, while for the others these are integrated
values. The values have been estimated from
two-dimensional Gaussian fits.
\end{flushleft}
\end{table}

\begin{figure}
\vbox{
    \psfig{file=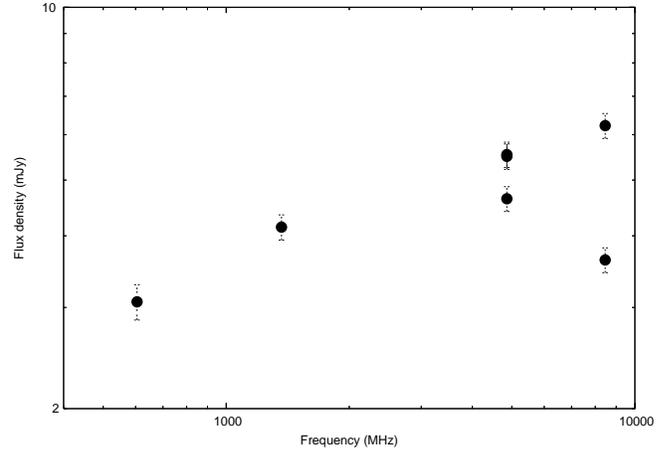,width=3.4in,angle=-90}
}
\caption[]{Spectrum of the core of J1453+3308.}
\end{figure}

\subsection{Polarisation properties}
As in Machalski et al. (2006) the preliminary polarisation properties of J1453+3308
are derived using the NVSS data (Condon et al. 1998) and our VLA CnD-array 4860 MHz
measurements.
The E-vectors superimposed on the total-intensity as well as polarised-intensity
images at 1400 and 4860 MHz are shown in Fig. 6.  The lobes of the source show significant
polarisation. The mean values of scalar polarisation at 4860 MHz for the northern and southern
lobes of the outer double are $\sim$8 and 20 per cent respectively. These values are
determined after clipping the images at 4 times the rms  noise and convolving the image with
the NVSS beam of 45$\times$45 arcsec$^2$. The average value for the inner double is $\sim$5
per cent. The average  scalar polarisation at 1400 MHz for the northern and southern lobes
of the outer double are $\sim$9 and 16 per cent respectively, while the corresponding value
for the inner double is $\sim$5 per cent. The northern component of the outer double shows no
evidence of  depolarisation while the southern lobe is depolarised by $\sim$20 per cent. 
The inner double shows no significant evidence of depolarisation. 

The rotation measure (RM) image of the source, obtained from the 1400-MHz and 4860-MHz  maps by
considering only those points which are 4 times above the rms noise, yields RM values
within $\sim$25 rad m$^{-2}$ (Fig. 7). Most regions of the source have an RM of only
a few rad m$^{-2}$, the average value for the source being $\sim$6 rad m$^{-2}$.
Although from two-frequency measurements it is not possible to resolve any n$\pi$ ambiguities in
the PA of the E-vectors, the low value of RM is consistent with known values for large radio sources
(e.g. Saikia \& Salter 1988 for a review). The Galactic longitude and latitude of the source are 
53.3$^\circ$ and  63.1$^\circ$ respectively. Considering the high Galactic latitude of the source, 
its RM is similar to other known sources in this direction suggesting that intrinsic contribution
to the RM is likely to be less than 20 rad m$^{-2}$. However these results are preliminary;
the RM distribution and the inferred magnetic field need to be determined from more
detailed multi-frequency observations.

The inferred magnetic field vectors obtained at present by  merely rotating the 
E-vectors at 4860 MHz by 90$^\circ$
shows that the field lines in the relaxed outer northern lobe are nearly circumferential in the
outer periphery. The field lines are roughly along the axis of the lobes for both the outer and
inner double, although the field lines tend to be orthogonal to the axis of the source in regions
of the outer double which are closest to the core or inner double. The field directions for the
northern and southern components of the inner double are $\sim$30 and 10$^\circ$ respectively.  
Some of the features appear somewhat unusual compared with the field distributions in other
radio galaxies (cf. Saripalli et al. 1996; Mack et al. 1997; Lara et al. 2000; Schoenmakers et al.
2000b). 
 
\begin{figure*}
\hbox{
   \psfig{file=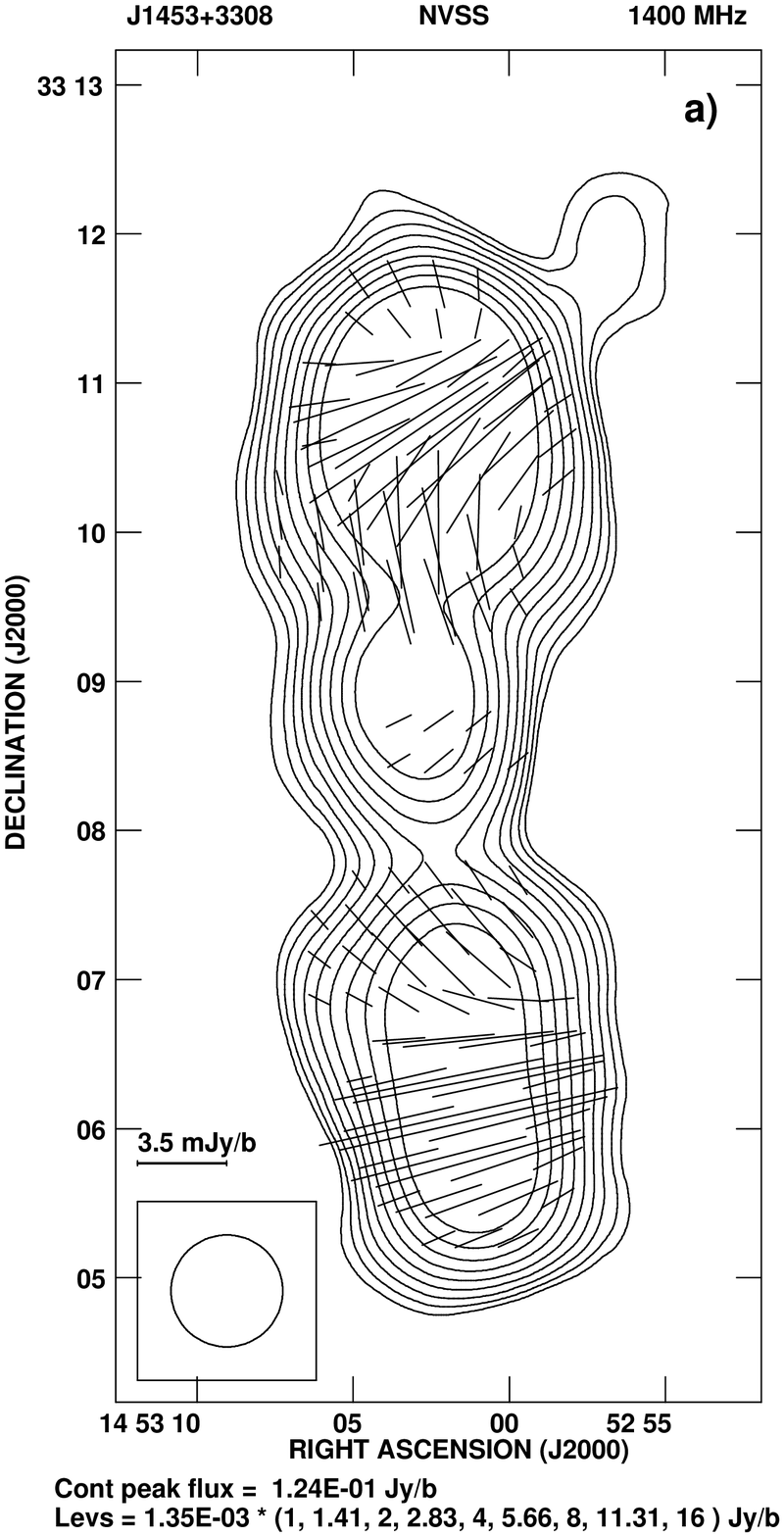,height=3.5in,angle=0}
   \psfig{file=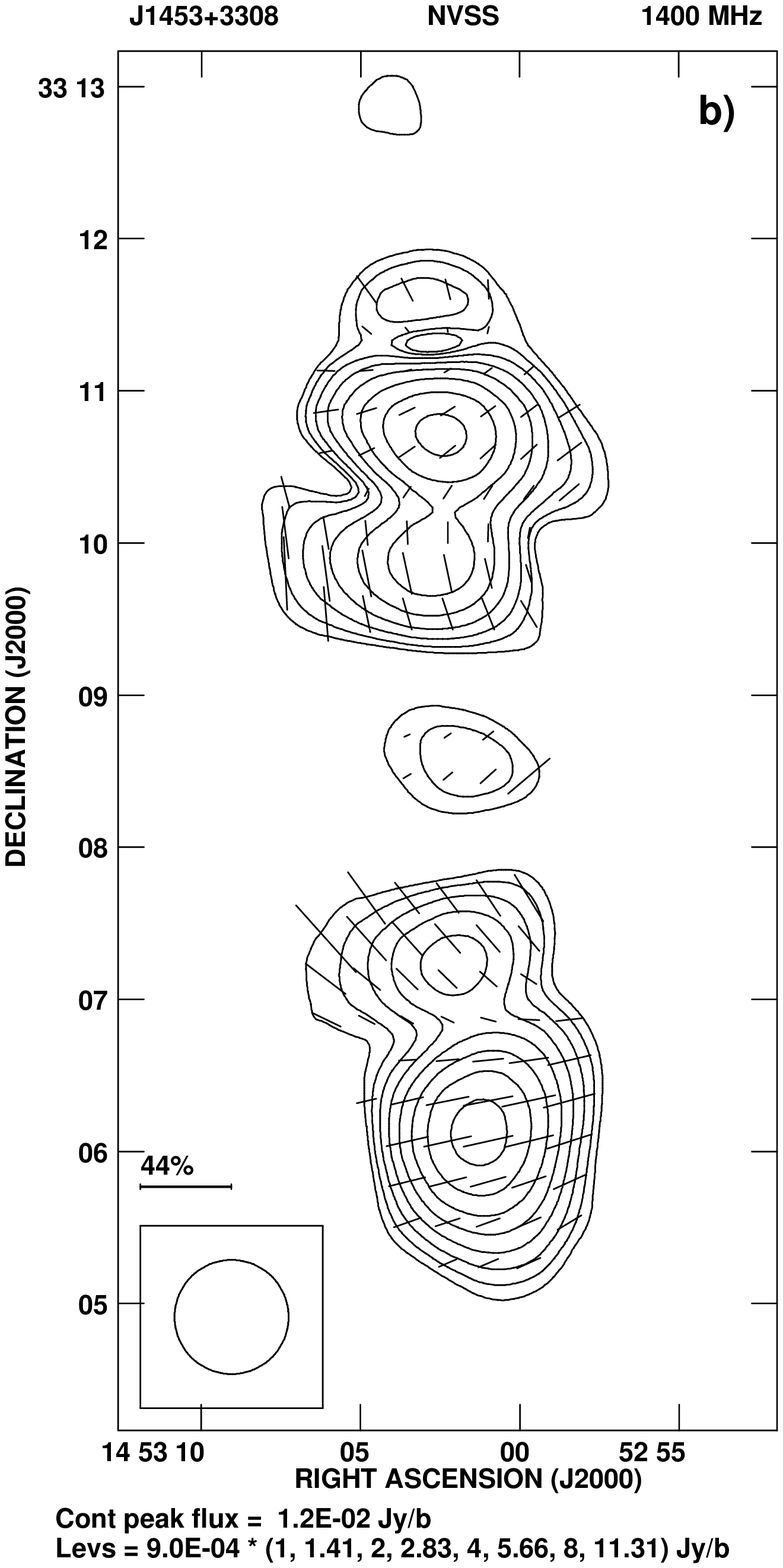,height=3.5in,angle=0}
   \psfig{file=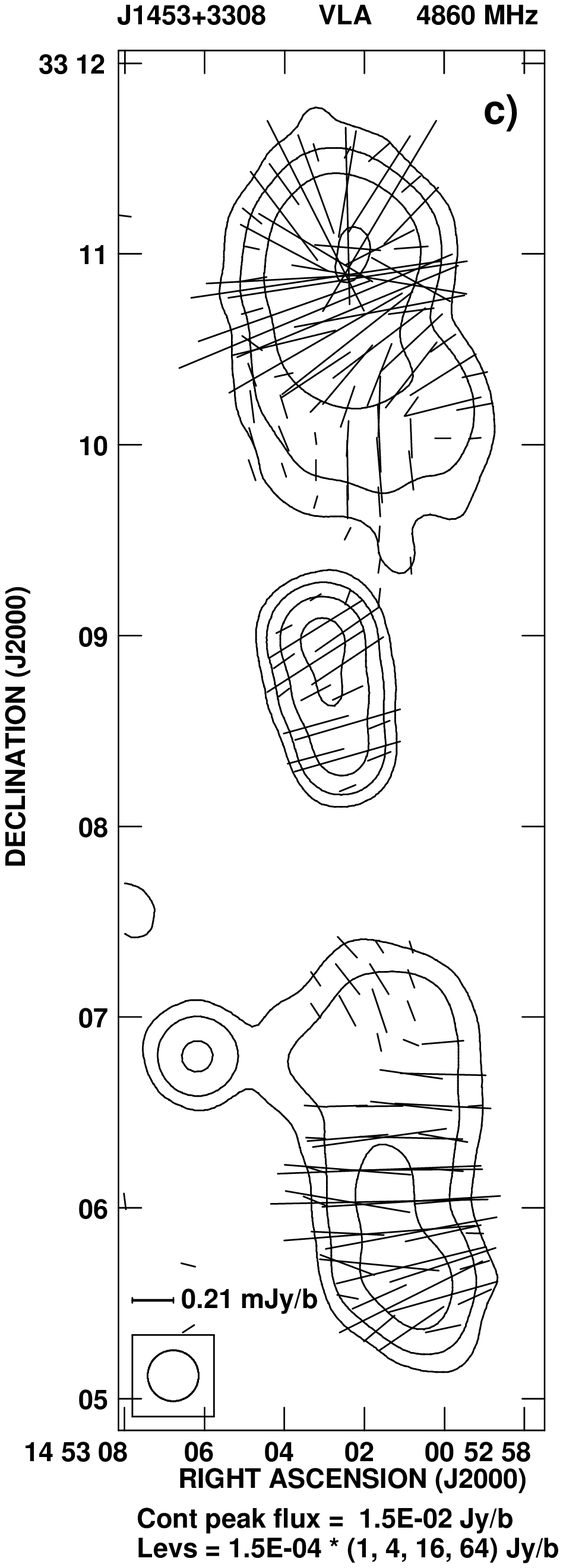,height=3.5in,angle=0}
   \psfig{file=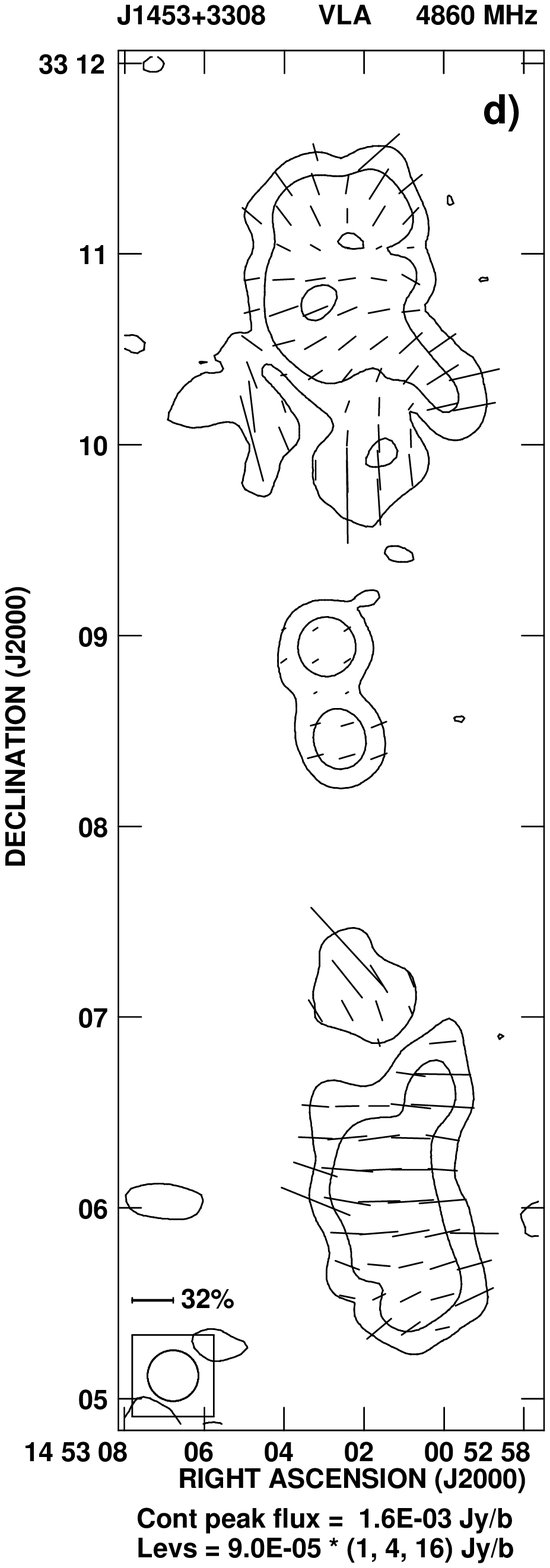,height=3.5in,angle=0}
}
\caption[]{
(a): The polarization E-vectors at 1400 MHz superimposed on the
total-intensity and (b) the corresponding fractional polarization vectors
superimposed on the polarized intensity contours. This is from the NVSS
and has an angular resolution of 45 arcsec. (c) The polarization E-vectors at 4860 MHz
superimposed on the total-intensity contours and (d) the
corresponding fractional polarization vectors superimposed on the polarized intensity contours.
}
\end{figure*}

\begin{figure}
\vbox{
    \psfig{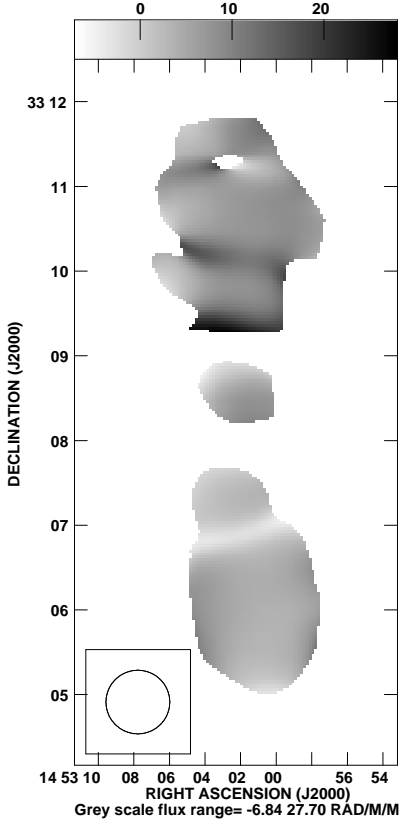}
}
\caption[]{Rotation measure distribution between 1400 and 4860 MHz with an angular resolution
of 45 arcsec. This is only indicative and requires confirmation from multi-frequency observations.}
\end{figure}

\section{Spectral ageing analysis}
In order to determine the spectral age in different parts of the lobes, i.e. the time
which elapsed since the radiating particles were last accelerated, we apply the standard
theory describing the time-evolution of the emission spectrum from particles with an
initial power-law energy distribution and distributed isotropically in pitch angle relative
to the magnetic field direction. The initial energy distribution corresponds to the initial 
(injection) spectral
index $\alpha_{\rm inj}$. The spectral break frequency above which the radio spectrum
steepens from the injected power law, $\nu_{\rm br}$, is related to the spectral age and the
magnetic field strength through

\[\tau_{\rm rad}=50.3\frac{B^{1/2}}{B^{2}+B^{2}_{\rm iC}}\{\nu_{\rm br}(1+z)\}^{-1/2} {\rm Myr},\]

\noindent
where $B_{\rm iC}$=0.318(1+$z$)$^{2}$ is the magnetic field strength equivalent to
the inverse-Compton microwave background radiation; $B$ and $B_{\rm iC}$ are expressed in units 
of nT, while $\nu_{\rm br}$ is in GHz.

\subsection{Determination of $\alpha_{\rm inj}$ and $\nu_{\rm br}$ values}
In order to determine a value of $\alpha_{\rm inj}$, we first fitted the CI (continuous
injection; Pacholczyk 1970) and JP (Jaffe \& Perola 1973) models of radiative losses to the 
flux densities of the outer lobes, and found that the uncertainties of the fitted 
$\alpha_{\rm inj}$ values are large. Also there is no evidence for significantly different 
values of this parameter in the opposite lobes. The typical sizes of 
hotspots are $\lapp$10 kpc (e.g. Jeyakumar \& Saikia 2000 and references therein) which
at the redshift of 0.249 for this source corresponds to an angular size of $\sim$3 arcsec.
This is similar to our GMRT full-resolution image at 1287 MHz which shows no significant
hotspots in either of the outer lobes. We have therefore preferred to use the JP rather
than the CI model for the outer lobes. Also the JP model gives an overall better fit to  
the spectra of the different strips discussed below.
The fit of the model to the flux densities of both the outer lobes together (column 9 in Table 3) is
shown in  Fig.~8a. This fit is used to estimate the  value of $\alpha_{\rm inj}$ for
the ageing analysis of the outer structure of the source. The corresponding fit, but with
the CI model applied to the flux densities of the inner double, is shown in Fig.~8b. It is worth
noting that both the fitted values of $\alpha_{\rm inj}$ are similar $\sim0.6$. 

Next, the total-intensity maps at the frequencies of 240, 334, 605, 1287 and 4860 MHz were
convolved to a common angular resolution of 15.2 arcsec.
Each lobe was then split into a number of strips, separated approximately by the resolution
element along the axis of the source, and the spectrum of each strip determined from 235
to 4860 MHz. Using the SYNAGE software (Murgia 1996) we searched for the best fit applying the
JP model to the spectra in the six strips covering the outer, northern lobe and the other
eight covering the outer, southern lobe. Although the resulting values of $\alpha_{\rm inj}$
for different strips show significant variations, the best fits with the same injection spectrum are 
achieved for $\alpha_{\rm inj}$=0.568. The results for a few typical strips
are shown in Fig. 9. In cases where the flux density of a particular strip at 235 MHz
appeared discrepant from the overall fit, we attempted to determine the values of
$\nu_{\rm br}$ with and without the discrepant point and found no significant difference.
The values of $\nu_{\rm br}$ including the 1$\sigma$ errors for the northern and southern
lobes are listed in Table 5.

\subsection{Magnetic field determination and radiative ages}
To determine the age of the particles in particular strips we have to estimate the magnetic
field strength in the corresponding strips. The values of the equipartition energy density and the 
corresponding magnetic field, B$_{\rm eq}$, are calculated using the revised formula given
by Beck \& Krause (2005). This formula (their equation (A18)) accounts for a ratio 
{\bf K}$_{0}$ between the number densities of protons and electrons in the energy range where
radiation losses are small. Assuming {\bf K}$_{0}$=70 implied by their equation (7) for
$\alpha\approx \alpha_{\rm inj}$=0.568, we obtain the revised magnetic field strength, $B_{\rm eq}$,
which depends on the low-frequency spectral index $\alpha$ in the observed synchrotron spectrum. The 
revised field, calculated assuming a filling factor of unity, $\alpha_{\rm inj}$=0.568, the 
flux densities at 334 MHz and a rectangular box for each strip
are listed in Table~5 which is arranged as follows. Column~1: identification of the strip; column 2:
the projected distance of the strip-centre from the radio core; column~3: the break frequency in GHz;
column~4: the reduced $\chi^{2}$ value of the fit; column~5: the revised magnetic field in nT;
column~6: the resulting synchrotron age of the particles in the given strip. These latter values
as a function of distance are plotted in Fig.~10. We have repeated the calculations using the 
classical equipartition magnetic field (e.g. Miley 1980) and have also
plotted the variation of age with distance using these fields. Within the uncertainties there is
no significant difference between the age estimates using the revised and classical 
equipartition magnetic fields.    

As expected the synchrotron age for both the outer lobes increases with distance
from the edges of the lobes. The maximum ages for the northern and southern lobes are
$\sim$47 and 58 Myr respectively. The northern lobe has a higher surface brightness
along most of its length, is closer to the nucleus and hence its younger spectral age may
be due to a combination of reacceleration of particles and better confinement.
A weighted least-squares fit
to the ages yields a mean separation velocity of the head from the radio-emitting plasma
of 0.036c for both the northern and southern lobes.
However, these separation velocities are referred to a region of the last
acceleration of emitting particles (shock at the end of the lobe) which advances from the
origin of the jets, i.e. the radio core, with an advance speed, $v_{\rm adv}$. If the backflow
has a speed of $v_{\rm bf}$, and  $v_{\rm bf}\approx v_{\rm adv}$, then an
average expansion speed is $\sim$0.018c. This would imply an age
of the outer structure of $\sim$96 and 134 Myr for the northern and southern lobes respectively,
significantly smaller than the $t_{\rm out}$ estimate of 215 Myr given by K2000.

It is worth noting that our derived age estimate for the inner double structure
is similar to that of K2000. Indeed, the fit with the CI model (cf. Fig.~8b) gives
$\alpha_{\rm inj}=0.566^{+0.052}_{-0.064}$ and $\nu_{\rm br}=856^{+1{\rm E}11}_{-848}$
GHz (the $\pm 1\sigma$ error is enormous due to the practically straight spectrum).
Our calculation of the magnetic field in the inner lobes gives $B_{\rm eq}=0.56\pm0.11$
nT which implies its spectral age to be $\sim$2 Myr and an apparent advance velocity of 
$\sim$0.1c for the lobe heads.
However these values have large uncertainties because the spectrum is practically straight.

\begin{figure}
\vbox{
   \psfig{file=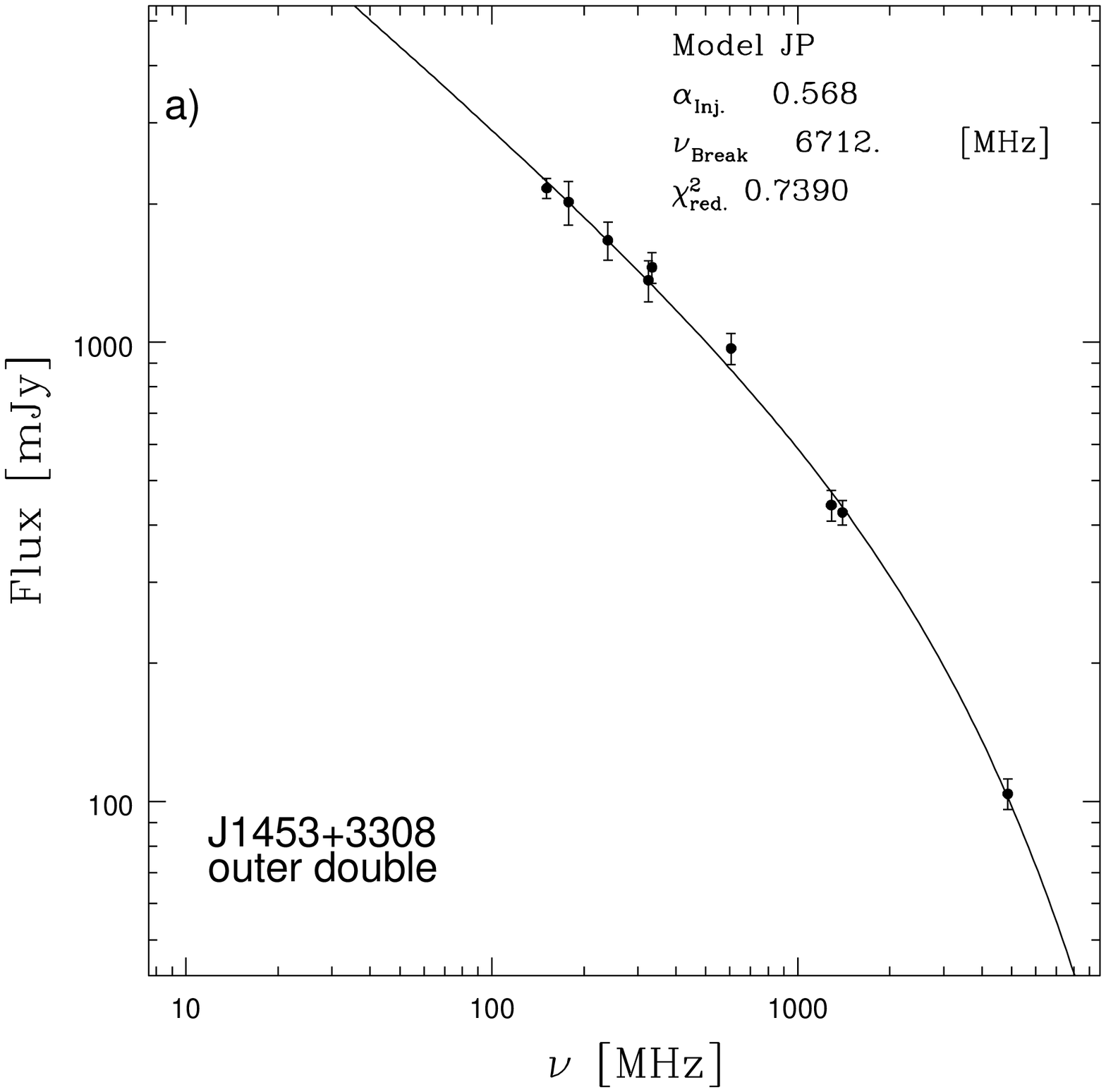,width=3.3in,angle=0}
   \psfig{file=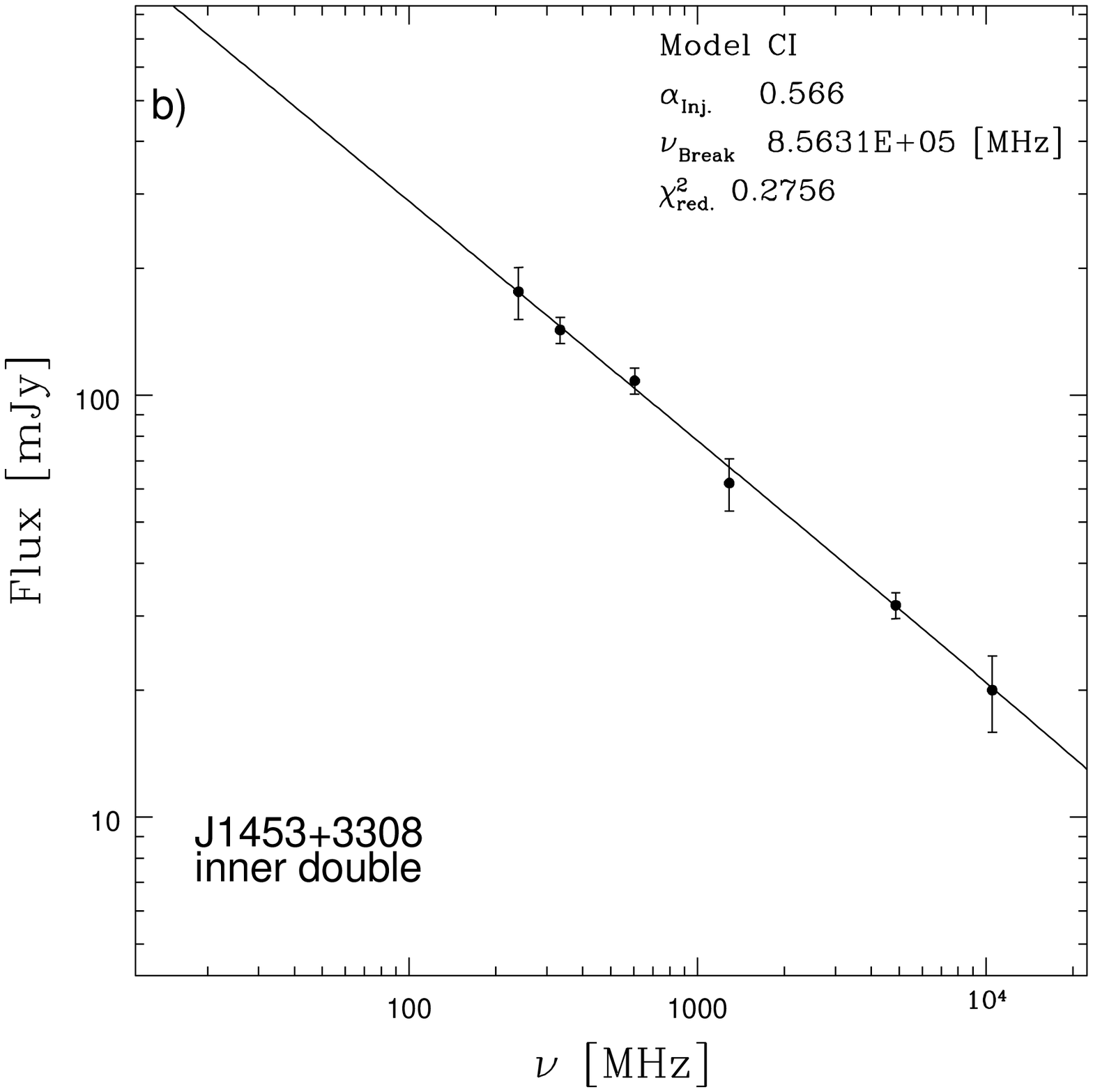,width=3.3in,angle=0}
}
\caption[]{Spectra of the outer and inner doubles fitted with the models of radiative
losses, as described in the text. Upper panel: the outer double fitted with the JP model;
lower panel: the inner double fitted with the CI model.}
\end{figure}

\begin{figure*}
\vbox{
   \hbox{
  \psfig{file=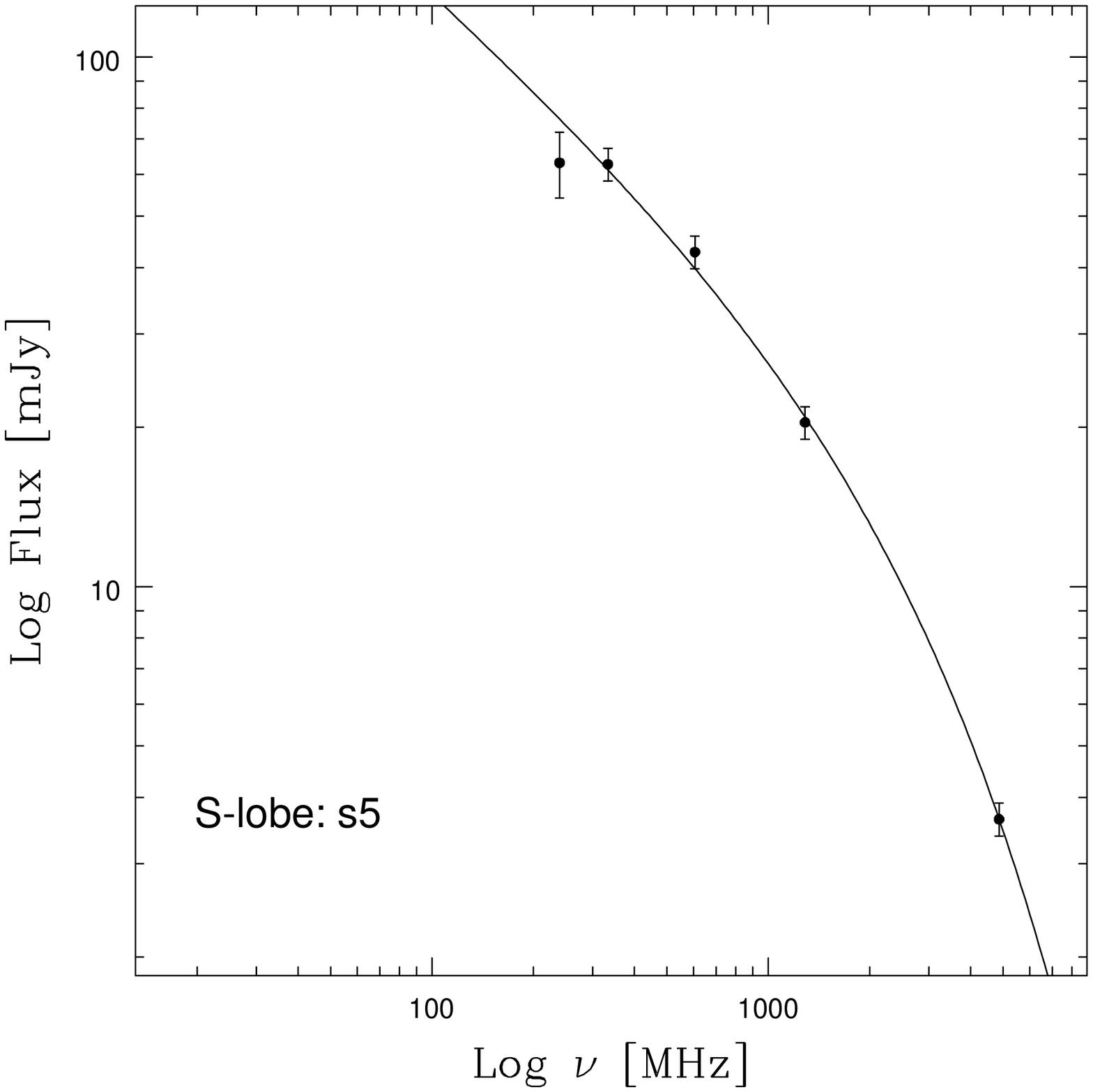,width=1.75in,angle=0}
  \psfig{file=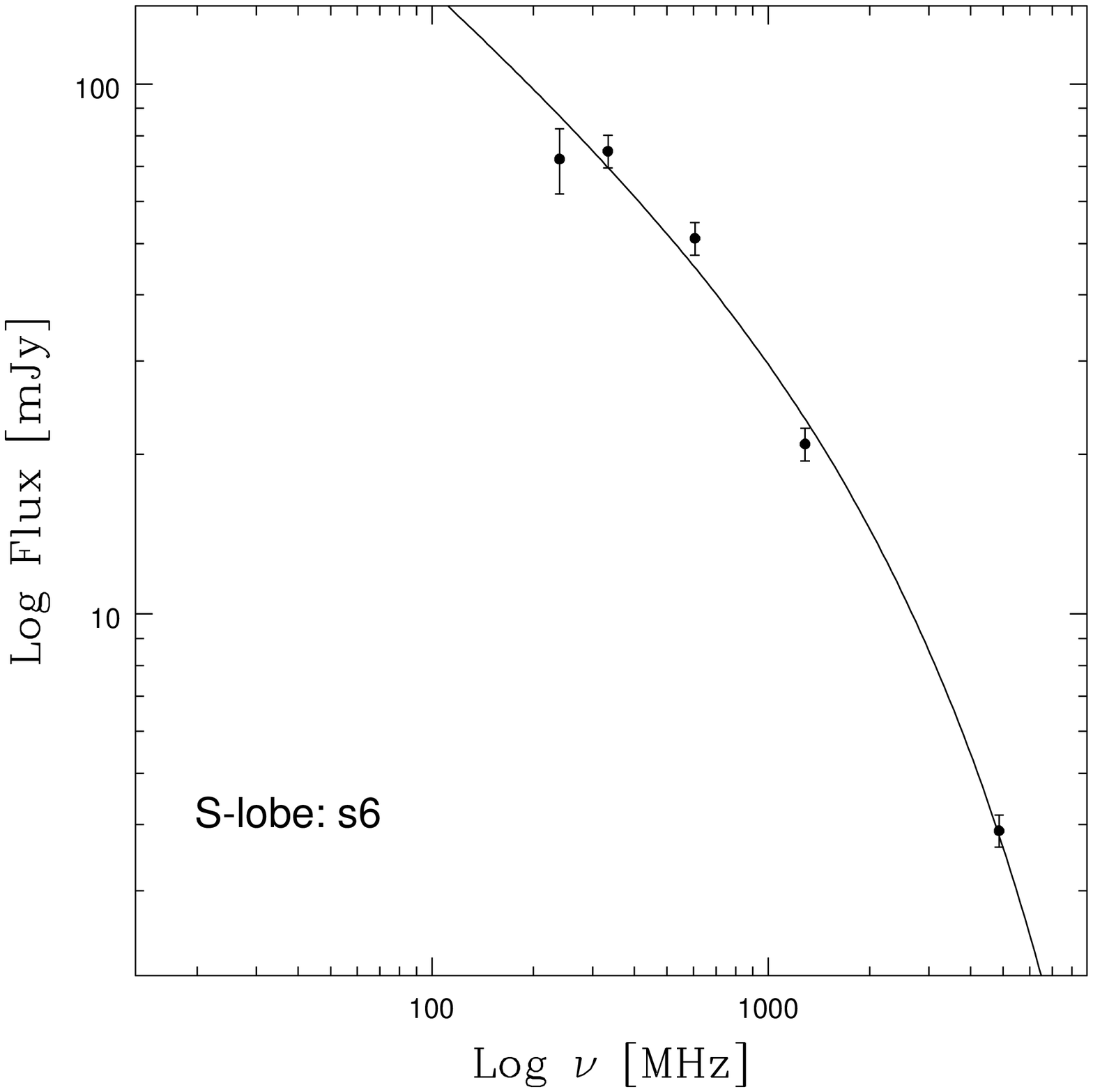,width=1.75in,angle=0}
  \psfig{file=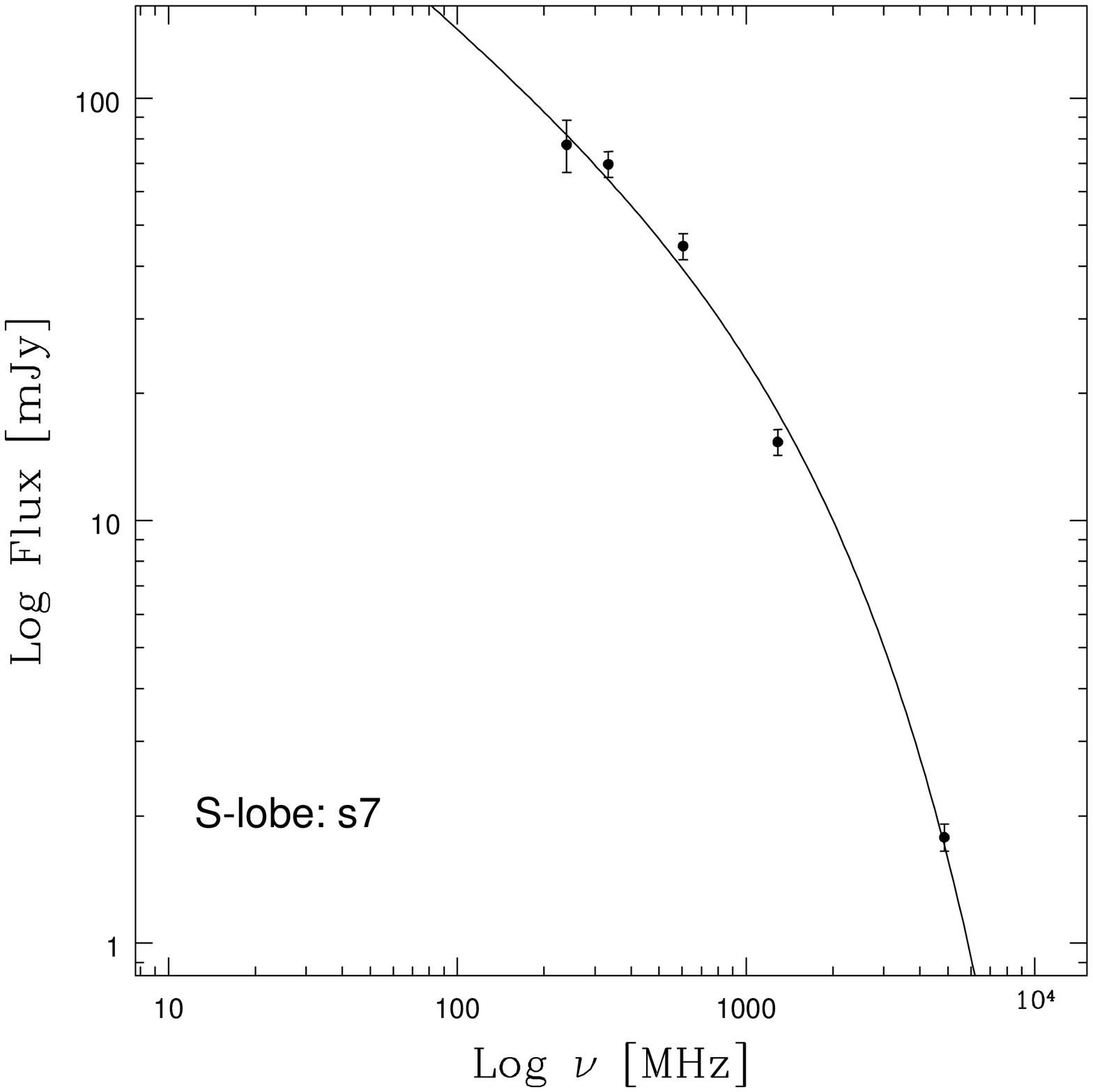,width=1.75in,angle=0}
  \psfig{file=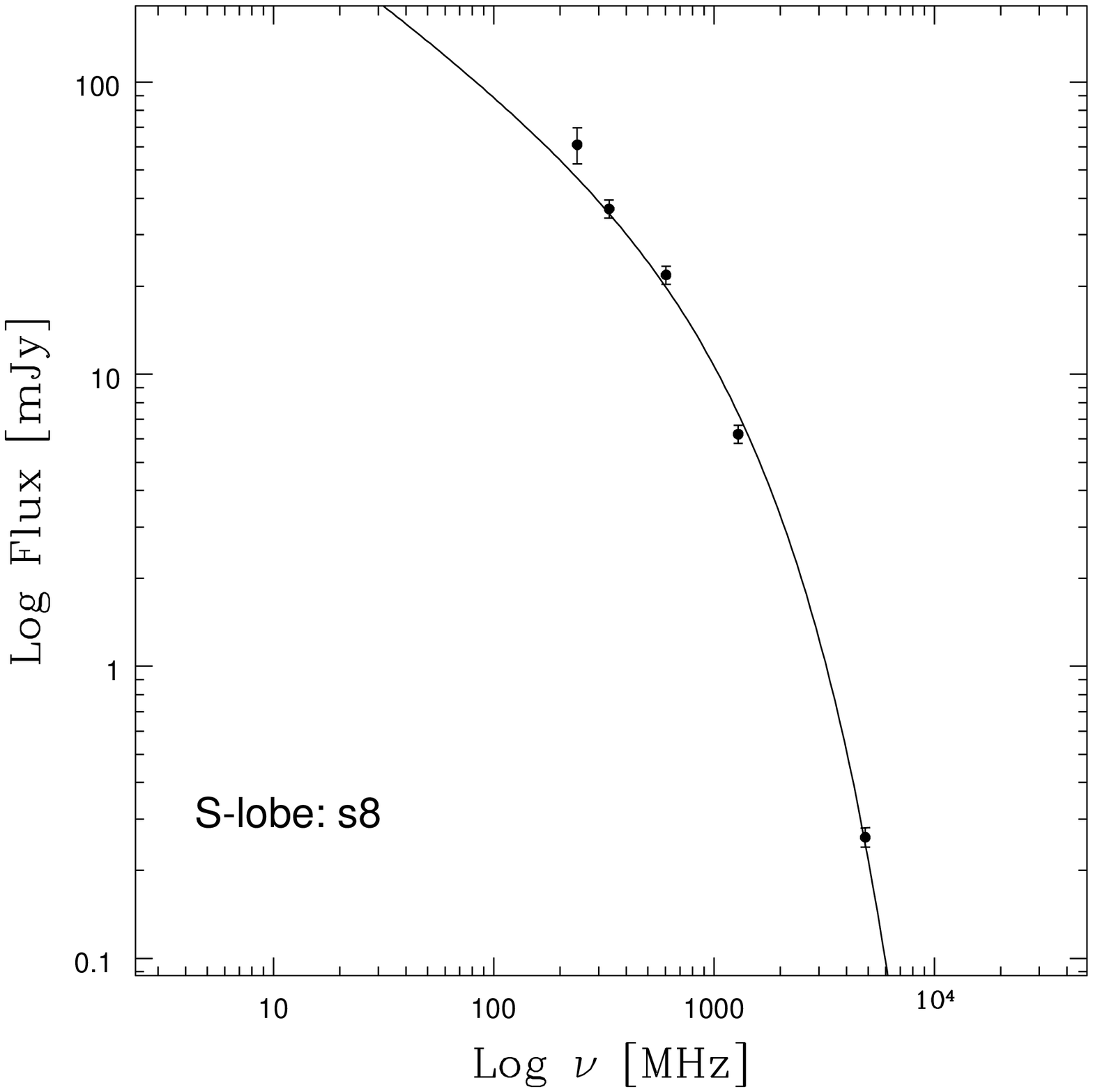,width=1.75in,angle=0}
    }
    \hbox{
  \psfig{file=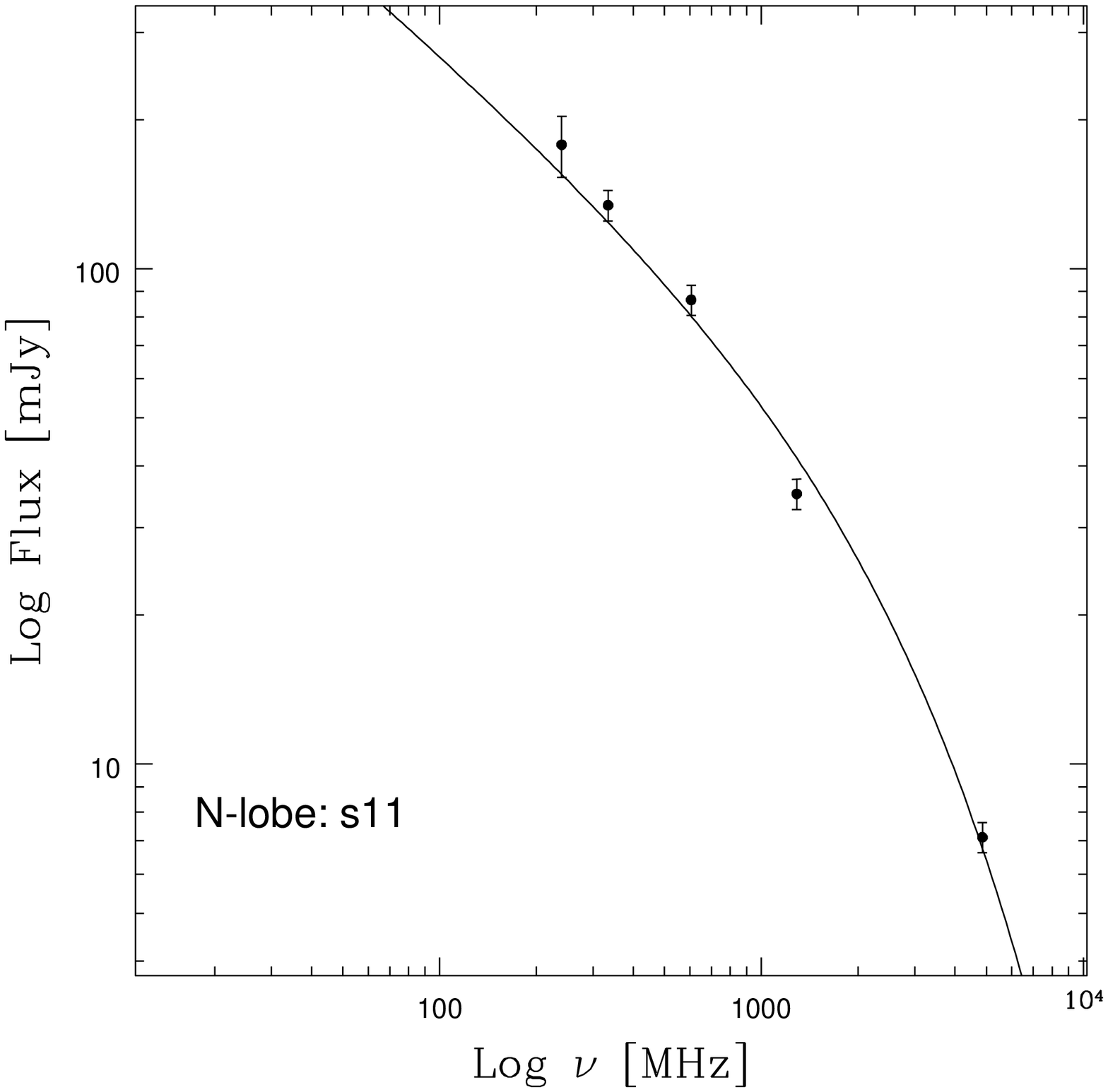,width=1.75in,angle=0}
  \psfig{file=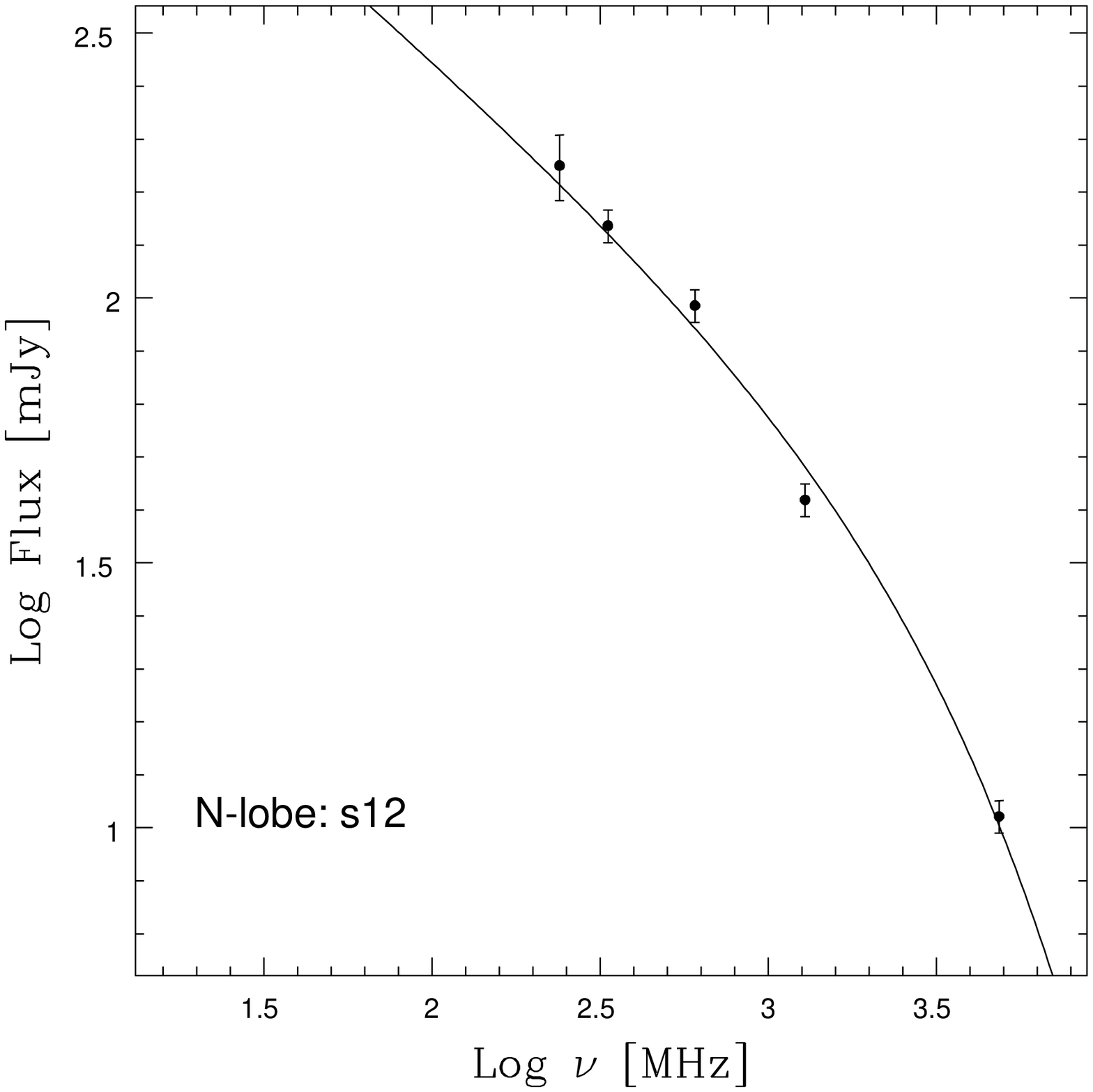,width=1.75in,angle=0}
  \psfig{file=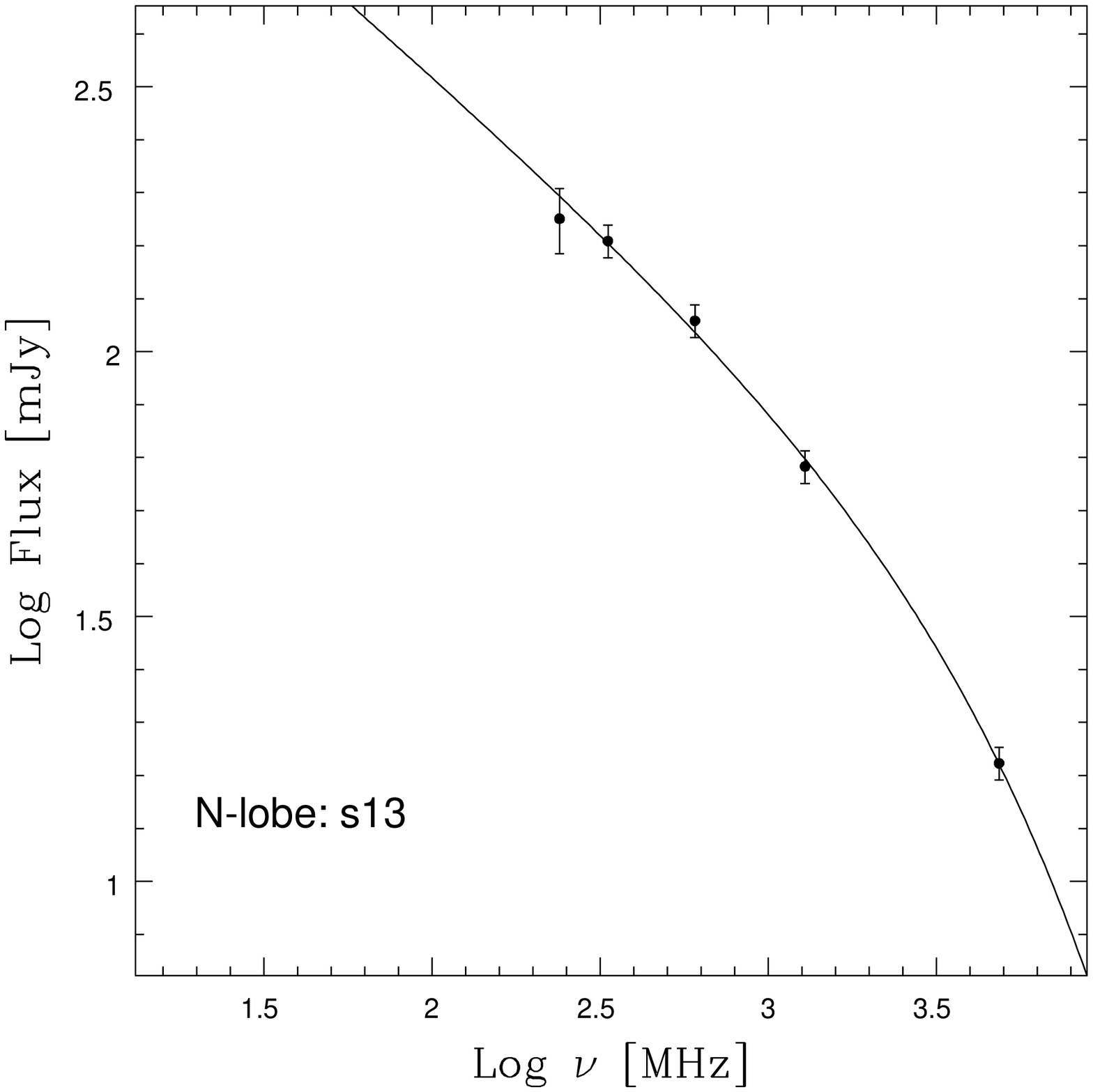,width=1.75in,angle=0}
  \psfig{file=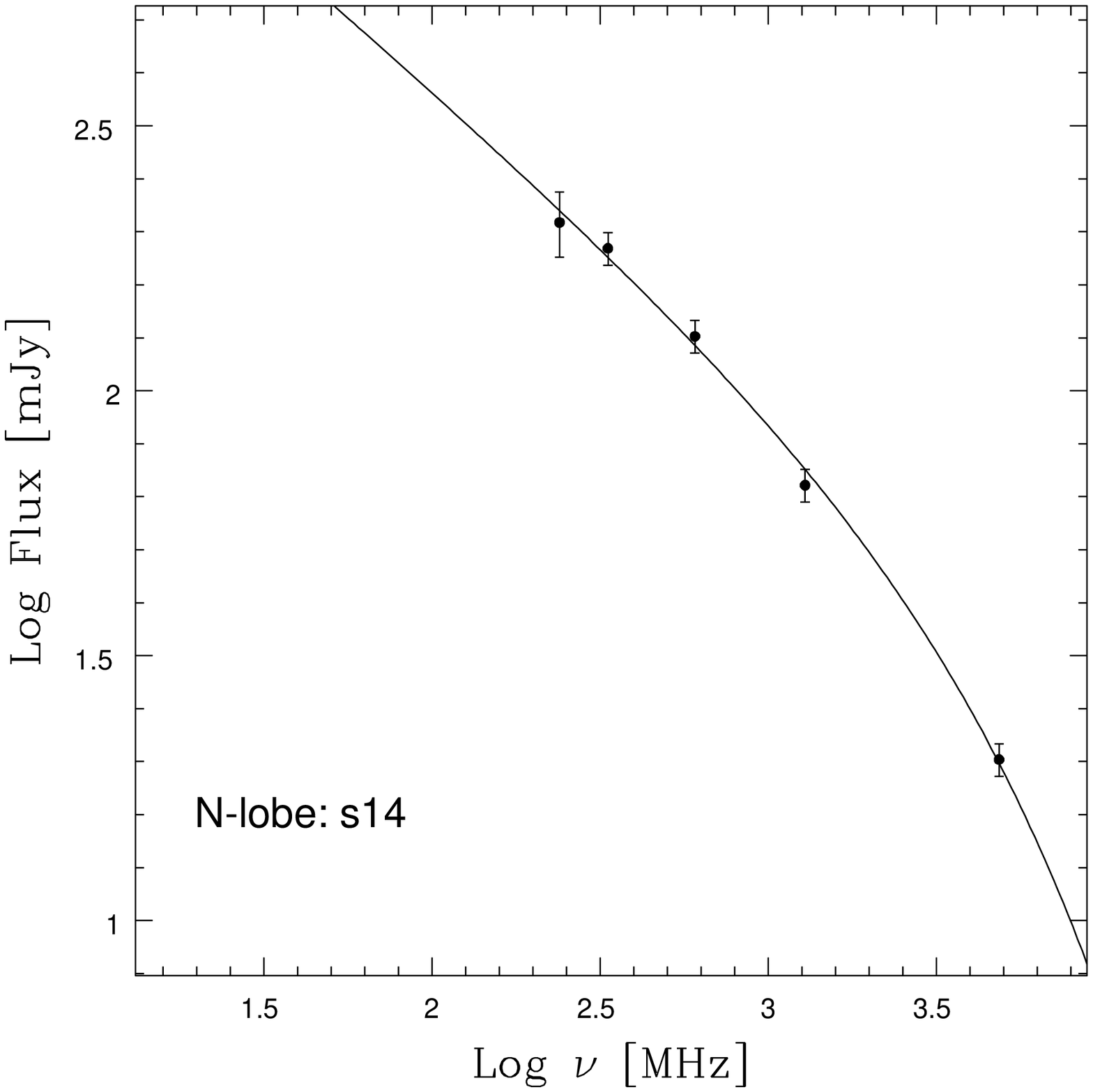,width=1.75in,angle=0}
   }
}
\caption[]{Typical spectra of the slices for the northern (upper panel) and southern (lower panel)
lobes of the outer double.  }
\end{figure*}

\begin{table}
\begin{center}
\caption{Results of JP model calculations with $\alpha_{\rm inj}$=0.568} 
\begin{tabular}{llllll}
\hline
Strip & Dist. & $\rm \nu_{br}$         & $\chi^{2}_{red}$ & $\rm B_{eq}(rev)$& $\rm \tau_{rad}$       \\
 & kpc   & GHz                   &                  & nT           & Myr                    \\
\hline
&&{\bf S-lobe} \\
s1 & 735 &$15^{+6570}_{-3}$    &4.45              & 0.35$\pm$0.07  &$18.7^{+2.3}_{-18}$    \\
s2 & 673 &$9.1^{+253}_{-1.3}$  &2.06              & 0.36$\pm$0.07  &$24.1^{+2.0}_{-9.9}$    \\
s3 & 611 &$6.9^{+16}_{-1.4}$   &1.27              & 0.35$\pm$0.07  &$27.7^{+3.2}_{-13}$    \\
s4 & 549 &$5.8^{+2.9}_{-1.7}$  &0.59              & 0.35$\pm$0.07  &$30.3^{+5.7}_{-5.5}$     \\
s5 & 487 &$4.9^{+2.6}_{-1.1}$  &0.33              & 0.36$\pm$0.07  &$32.8^{+4.4}_{-6.3}$     \\
s6 & 425 &$4.6^{+21}_{-0.4}$   &2.53              & 0.38$\pm$0.07  &$33.7^{+1.4}_{-19}$     \\
s7 & 363 &$2.9^{+4.1}_{-0.2}$  &3.18              & 0.37$\pm$0.07  &$42.7^{+1.2}_{-15}$    \\
s8 & 301 &$1.6^{+1.6}_{-0.1}$  &3.15              & 0.31$\pm$0.06  &$58.2^{+0.9}_{-17}$    \\
     &   &                     &                  &              &                        \\

&&{\bf N-lobe} \\
s9 & 244  &$2.5^{+9.7}_{-0.1}$ &17.9              & 0.32$\pm$0.06  &$46.6^{+1.4}_{-25}$  \\
s10& 286  &$3.8^{+4.0}_{-0.3}$ &3.69              & 0.37$\pm$0.07  &$37.3^{+1.4}_{-11}$  \\
s11& 348  &$4.7^{+349}_{-0.4}$ &2.79              & 0.40$\pm$0.08  &$32.6^{+1.4}_{-29}$  \\
s12& 410  &$6.6^{+43}_{-0.7}$  &2.04              & 0.41$\pm$0.08  &$27.4^{+1.6}_{-17}$  \\
s13& 472  &$10^{+6.3}_{-4.6}$  &0.46              & 0.42$\pm$0.08  &$21.8^{+7.5}_{-4.7}$  \\
s14& 534  &$11^{+36}_{-3.3}$   &0.47              & 0.44$\pm$0.08  &$20.3^{+3.8}_{-10}$ \\
     &    &                    &                  &                &                      \\
\hline
\end{tabular}
\end{center}
\end{table}

Nevertheless, while interpreting these numbers caveats related to the evolution of
the local magnetic field in the lobes need to be borne in mind (e.g. Rudnick, Katz-Stone
\& Anderson 1994; Jones, Ryu \& Engel 1999; Blundell \& Rawlings 2000).
While K2000 have suggested that spectral and dynamical ages are comparable if
bulk backflow and both radiative and adiabatic losses are taken into account in a
self-consistent manner, Blundell \& Rawlings (2000) suggest that this may be so only
in the young sources with ages much less than 10 Myr. In the study of the FRII type
giant radio galaxy, J1343+3758, Jamrozy et al. (2005) find the dynamical age to be
approximately 4 times the maximum synchrotron age of the emitting particles.

\begin{figure}
\vbox{
    \psfig{file=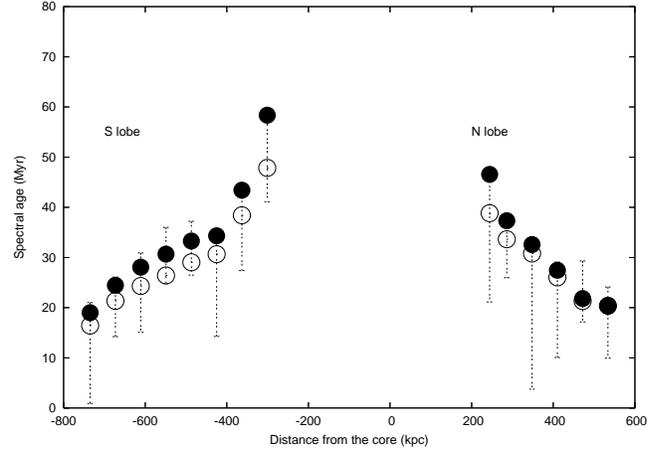,width=3.4in,angle=-90}
}
\caption[]{Radiative age of the relativistic particles in the outer lobes of J1453+3308
plotted against the distance from the radio core using the revised (filled circles) and
classical (open circles) equipartition magnetic fields. }
\end{figure}

\section{Concluding remarks}
We present the results of multifrequency radio observations of the
double-double radio galaxy (DDRG), J1453+3308, using both the GMRT and the VLA. 

\begin{enumerate}
\item The images show more details of the well-known pair of doubles which
are misaligned by $\sim$7.5$^\circ$. The symmetry parameters of the outer- and
inner-double lobes suggest intrinsic asymmetries over these length scales. The
luminosity of the outer double is in the FRII category although it has no prominent
hotspot, while the luminosity of the inner double is below the dividing line although 
it has an edge-brightened structure characteristic of FRII sources.   

\item The radio core has a mildly inverted spectrum, with some evidence of variability.
The weak, compact source close to the southern lobe of the outer double is
unrelated.

\item The magnetic field lines exhibit significant structure, with the lines being 
orthogonal to the source axis in the outer lobes closest to the radio core. The rotation
measure is only a few rad m$^{-2}$ over most of the source, with the maximum value being
$\sim$25 rad m$^{-2}$.  These results are tentative since they are based on observations
at only two frequencies and require confirmation from multi-frequency observations.

\item The integrated spectrum is curved with evidence of significant flattening 
below $\sim$300 MHz. Using our images from 235 to 4860 MHz smoothed to a similar resolution
of  $\sim$15 arcsec,  we fit the observed spectra of the outer and inner double structures
with the models of radiative losses, and we determine the characteristic frequency breaks
in fourteen strips cut through the outer lobes and transverse to the source axis. The best
fits for the outer lobes are obtained with the JP model and the injection spectrum 
$\alpha_{\rm inj}$=0.568, while the best fit to the spectrum of the inner double is obtained 
with the CI model and $\alpha_{\rm inj}$=0.566. The injection spectra for both the outer and
inner doubles are similar within the uncertainties.

\item The synchrotron age of emitting particles in both the lobes of the outer double
increases with distance from the edges of the lobes. The maximum ages for the northern and
southern lobes are $\sim$47 Myr and $\sim$58 Myr respectively.
The dependences of the radiative age vs. distance from the core imply a mean separation
velocity of the lobe's head from the radio-emitting plasma of 0.036c for both
the northern and southern lobes. However, assuming presence of a backflow with its
backward speed comparable to the advance speed of the (jet) head, an average advance speed would be
about 0.018c which yields a maximum age of $\sim$134 Myr which is significantly smaller than the
value suggested by K2000.

\item The spectral age of $\sim$2 Myr obtained for the inner double is similar to that
estimated by K2000. The value of 2 Myr implies an apparent advance velocity of $\sim$0.1c 
for the lobe heads of the inner structure.
However these values have large uncertainties because the spectrum is practically straight.

\end{enumerate}

\section*{Acknowledgments} 
We thank an anonymous referee for his helpful and detailed comments on the manuscript, 
Bill Cotton and Ger de Bruyn for clarifying the VLSS and WENSS flux density scales,
our colleagues for their comments and the staffs of GMRT and VLA for their help with the
observations. The Giant Metrewave Radio Telescope is a national facility operated by the
National Centre for Radio Astrophysics of the Tata Institute of Fundamental Research. 
The National Radio Astronomy Observatory  is a facility of the National Science Foundation
operated under co-operative agreement by Associated Universities Inc. 
This research has made use of the NASA/IPAC extragalactic database (NED) which is operated
by the Jet Propulsion Laboratory, Caltech, under contract with the National Aeronautics
and Space Administration. MJ acknowledges the access to the SYNAGE software kindly provided 
by Dr K.-H. Mack and Dr M. Murgia (Istituto di Radioastronomia, Bologna, Italy). CK and DJS
also thank Dr M. Murgia for access to the software and useful discussions.
JM and MJ were partly supported by the Polish State funds for scientific research in years
2005--2007 under contract No. 0425/PO3/2005/29.

{}


\begin{thebibliography}{}
\bibitem[]{}  Baars J.W.M., Genzel R., Pauliny-Toth I.I.K., Witzel A. 1977, A\&A, 61, 99
\bibitem[]{}  Beck R., Krause M., 2005, Astron. Nach., 326, 414
\bibitem[]{}  Blundell K.M., Rawlings S., 2000, AJ, 119, 1111
\bibitem[]{}  Colla G., et al., 1970, A\&AS, 1, 281
\bibitem[]{}  Condon J.J., Cotton W.D., Greisen E.W., Yin Q.F., Perley R.A., Taylor G.B., Broderick J.J., 
              1998, AJ, 115, 1693
\bibitem[]{}  Jaffe W.J., Perola G.C., 1973, A\&A, 26, 423
\bibitem[]{}  Jamrozy M., Machalski J., Mack K.-H., Klein U., 2005, A\&A, 433, 467 	
\bibitem[]{}  Jeyakumar S., Saikia D.J., 2000, MNRAS, 311, 397
\bibitem[]{}  Jones T.W., Ryu D., Engel A., 1999, ApJ, 512, 105
\bibitem[]{}  Kaiser C.R., Schoenmakers A.P., R\"{o}ttgering H.J.A., 2000, MNRAS, 315, 381 (K2000)
\bibitem[]{}  Lara L., M\'arquez I., Cotton W.D., Feretti L., Giovannini G., Marcaide J.M., Venturi T.,
              1999, A\&A, 348, 699
\bibitem[]{}  Lara L., Mack K.-H., Lacy M., Klein U., Cotton W.D., Feretti L., Giovannini G., Murgia M., 2000, 
              A\&A, 356, 63
\bibitem[]{}  Machalski J., Jamrozy M., Zola S., Koziel D., 2006, A\&A, 454, 85 
\bibitem[]{}  Mack K.-H., Klein U., O'Dea C.P., Willis A.G., 1997, A\&AS, 123, 423
\bibitem[]{}  Miley G.K., 1980, ARA\&A, 18, 165	
\bibitem[]{}  Murgia M., 1996, Laurea Thesis, University of Bologna 
\bibitem[]{}  Pacholczyk A.G., 1970, Radio Astrophysics, W.H. Freeman, San Francisco 
\bibitem[]{}  Pilkington J.D.H., Scott P.F., 1965, MemRAS, 69, 183 
\bibitem[]{}  Riley J.M.W., Waldram E.M., Riley J.M., 1999, MNRAS, 306, 31
\bibitem[]{}  Rudnick L., Katz-Stone D.M., Anderson M.C., 1994, ApJS, 90, 955
\bibitem[]{}  Saikia D.J., Salter C.J., 1988, ARA\&A, 26, 93
\bibitem[]{}  Saikia D.J., Konar C., Kulkarni V.K., 2006, MNRAS, 366, 1391
\bibitem[]{}  Saripalli L., Mack K.-H., Klein U., Strom R., Singal A.K., 1996, A\&A, 306, 708
\bibitem[]{}  Schoenmakers A.P., 1999, PhD thesis, Universiteit Utrecht
\bibitem[]{}  Schoenmakers A.P., de Bruyn A.G., R\"ottgering H.J.A., van der Laan H., Kaiser C.R., 2000a,	
              MNRAS, 315, 371 (S2000a)
\bibitem[]{}  Schoenmakers A.P., Mack K.-H., de Bruyn A.G., R\"{o}ttgering H.J.A., Klein U., 
              van der Laan H., 2000b, A\&AS, 146, 293
\bibitem[]{}  Subrahmanyan R., Saripalli L., Hunstead R.W., 1996, MNRAS, 279, 257
\bibitem[]{}  Spergel D.N. et al., 2003, ApJS, 148, 175
\end{thebibliography}
\end{document}